\RequirePackage[l2tabu, orthodox]{nag}
\documentclass[conference,compsoc]{IEEEtran}

\usepackage[T1]{fontenc}
\usepackage[super]{nth}
\usepackage{romannum}
\usepackage[utf8]{inputenc}
\usepackage[english]{babel}
\usepackage{threeparttable} 
\usepackage[mode=text, detect-all, binary-units, range-units=single,
per-mode=symbol]{siunitx}
\AtBeginDocument{
\DeclareSIUnit\Byte{byte}
\DeclareSIUnit\bit{b}
\DeclareSIUnit\kilo{K}
\DeclareSIUnit\Mb{\mega\bit}
\DeclareSIUnit\Gb{\giga\bit}
\DeclareSIUnit\KB{\kilo\byte}
\DeclareSIUnit\MB{\mega\byte}
\DeclareSIUnit\GB{\giga\byte}
\DeclareSIUnit\kbps{\kb\per\second}
\DeclareSIUnit\Mbps{\Mb\per\second}
\DeclareSIUnit\Gbps{\Gb\per\second}
\DeclareSIUnit\ns{\nano\second}
}

\usepackage[backend=biber, style=numeric,giveninits=false,maxnames=99]{biblatex}
\usepackage[textsize=tiny]{todonotes}
\usepackage{setspace} 
\usepackage{wrapfig}
\usepackage{microtype}
\usepackage{url}
\usepackage{graphicx}
\usepackage{amsmath}
\usepackage{bookmark}
\usepackage{xcolor}
\usepackage{acronym}
\usepackage{subcaption}
\usepackage{tabularx}
\usepackage[inline, shortlabels]{enumitem}
\usepackage{multirow}
\usepackage{listings}
\usepackage{xspace}
\usepackage{booktabs}
\usepackage{soul}
\usepackage{tablefootnote}
\usepackage{pifont}
\usepackage[autostyle=true,english=american]{csquotes}
\AtBeginDocument{

}

\makeatletter
\patchcmd{\@setref}{\bfseries??}{\colorbox{red!30}{\detokenize{#3}}}{}{}
\def\abx@missing#1{\colorbox{red!30}{\detokenize{#1}}}
\makeatother

\AtEveryBibitem{\clearfield{isbn}}
\AtEveryBibitem{\clearfield{issn}}
\AtEveryBibitem{\clearlist{institution}}
\AtEveryBibitem{\clearfield{doi}}
\AtEveryBibitem{\clearlist{publisher}}
\AtEveryBibitem{\clearlist{location}}
\AtEveryBibitem{\clearfield{series}}
\AtEveryBibitem{\clearfield{month}}
\AtEveryBibitem{\clearfield{note}}
\AtEveryBibitem{\clearfield{pages}}
\AtEveryBibitem{\clearfield{pagetotal}}
\AtEveryBibitem{\clearfield{volume}}
\AtEveryBibitem{\clearfield{location}}
\AtEveryBibitem{\clearfield{address}}
\AtEveryBibitem{
	\ifboolexpr{test {\ifentrytype{article}}}{\clearfield{number}}{}
        \ifboolexpr{not test {\ifentrytype{misc}}}{\clearfield{url}}{}
}
\DeclareSourcemap{
        \maps[datatype=bibtex]{
                \map{
                        \step[fieldsource=booktitle,
                                match={.*Internet Measurement Conference.*},
                                replace={ACM IMC}]
                        \step[fieldsource=booktitle,
                                match={.*Network Traffic Measurement and Analysis Conference.*},
                                replace={IEEE TMA}]
                        \step[fieldsource=booktitle,
                                match={.*Workshop on the Evolution, Performance,
                                and Interoperability of QUIC.*},
                                replace={ACM EPIQ}]
                        \step[fieldsource=booktitle,
                                match={.*Networked Systems Design and Implementation.*},
                                replace={USENIX NSDI}]
                        \step[fieldsource=booktitle,
                                match={.*Hot Topics in Storage and File Systems.*},
                                replace={USENIX HotStorage}]
                        \step[fieldsource=booktitle,
                                match={.*Hot Topics in Networks.*},
                                replace={ACM HotNets}]
                        \step[fieldsource=booktitle,
                                match={.*Hot Topics in Operating Systems.*},
                                replace={ACM HotOS}]
                        \step[fieldsource=booktitle,
                                match={.*Hot Topics in Cloud Computing.*},
                                replace={USENIX HotCloud}]
                        \step[fieldsource=booktitle,
                                match={.*sigmm workshop on network and operating systems support for digital audio and video.*},
                                replace={ACM NOSSDOV}]
                        \step[fieldsource=booktitle,
                                match={.*Operating Systems Design and Implementation.*},
                                replace={USENIX OSDI}]
                        \step[fieldsource=booktitle,
                                match={.*Annual Technical Conference.*},
                                replace={USENIX ATC}]
                        \step[fieldsource=booktitle,
                                match={.*Conference on File and Storage Technologies.*},
                                replace={USENIX FAST}]
                        \step[fieldsource=booktitle,
                                match={.*International Conference on Measurement and Modeling of Computer Systems.*},
                                replace={ACM SIGMETRICS}]
                        \step[fieldsource=booktitle,
                                match={.*Symposium on Cloud Computing.*},
                                replace={ACM SoCC}]
                        \step[fieldsource=booktitle,
                                match={.*International Symposium on Field-Programmable Custom Computing Machines.*},
                                replace={IEEE FCCM}]
                        \step[fieldsource=booktitle,
                                match={.*International Symposium on Microarchitecture.*},
                                replace={IEEE/ACM MICRO}]
                        \step[fieldsource=booktitle,
                                match={.*EuroSys Conference.*},
                                replace={ACM EuroSys}]
                        \step[fieldsource=booktitle,
                                match={.*Web Conference.*},
                                replace={ACM WWW}]
                        \step[fieldsource=booktitle,
                                match={.*European Conference on Computer Systems.*},
                                replace={ACM EuroSys}]
                        \step[fieldsource=booktitle,
                                match={.*Symposium on Operating Systems Principles.*},
                                replace={ACM SOSP}]
                        \step[fieldsource=booktitle,
                                match={.*(USENIX Security Symposium|USENIX Conference on Security).*},
                                replace={USENIX Security}]
                        \step[fieldsource=booktitle,
                                match={.*(ACM SIGSAC Conference on Computer and Communications Security).*},
                                replace={ACM CCS}]
                        \step[fieldsource=booktitle,
                                match={.*USENIX Symposium on Internet Technologies and Systems.*},
                                replace={USENIX USITS}]
                        \step[fieldsource=booktitle,
                                match={.*Emerging Networking E.periments and Technologies.*},
                                replace={ACM CoNEXT}]
                        \step[fieldsource=booktitle,
                                match={.*emerging Networking E.periments and Technologies.*},
                                replace={ACM CoNEXT}]
                        \step[fieldsource=booktitle,
                                match={.*Asia-Pacific Workshop on Networking.*},
                                replace={ACM APNet}]
                        \step[fieldsource=booktitle,
                                match={.*Asia-Pacific Workshop on Systems.*},
                                replace={ACM APSys}]
                        \step[fieldsource=booktitle,
                                match={.*Roedunet.*},
                                replace={IEEE RoEduNet}]
                        \step[fieldsource=booktitle,
                                match={.*Architectural Support for Programming Languages and Operating Systems.*},
                                replace={ACM ASPLOS}]
                        \step[fieldsource=booktitle,
                                match={.*Symposium on Software Defined Networking Research.*},
                                replace={ACM SOSR}]
                        \step[fieldsource=booktitle,
                                match={.*Architectures? for Networking and Communications Systems.*},
                                replace={ACM ANCS}]
                        \step[fieldsource=booktitle,
                                match={.*SIGMOD International Conference on Management of Data.*},
                                replace={ACM SIGMOD}]
                        \step[fieldsource=journal,
                                match={.*Proceedings on Privacy Enhancing
                                Technologies.*},
                                replace={PETS}]
                        \step[fieldsource=booktitle,
                                match={.*(SIGCOMM|Special Interest Group on Data Communication|Applications, Technologies, Architectures, and Protocols|Communications Architectures and Protocols).*},
                                replace={ACM SIGCOMM}]
                        \step[fieldsource=journal,
                                match={.*IEEE Journal on Selected Areas in Communication.*},
                                replace={IEEE JSAC}]
                        \step[fieldsource=journal,
                                match={.*Comput. Commun. Rev..*},
                                replace={ACM SIGCOMM CCR}]
                }
        }
}
\DefineBibliographyStrings{english}{
        in               = {},
        page             = {p\adddot},
        pages            = {pp\adddot},
        june             = {Jun\adddot},
        july             = {Jul\adddot},
        september        = {Sep\adddot},
}

\hypersetup{pdffitwindow=true,pdfpagelayout=SinglePage,pdfpagemode=UseThumbs}

\newcommand{\homa}{{Homa}\xspace}
\newcommand{\khoma}{{Homa/Linux}\xspace}

\newcommand{\sdp}{SMT\xspace}
\newcommand{\ktls}{{kTLS}\xspace}
\newcommand{\tcpcrypt}{{TcpCrypt}\xspace}

\newcommand{\ebpf}{{eBPF}\xspace}
\newcommand{\rktio}{{rkt-io}\xspace}

\newcommand{\utcp}{{$\mathrm{\text\textmu}$TCP}}
\newcommand{\minion}{{Minion}\xspace}
\newcommand{\bh}[1]{{\noindent{\textbf{#1.}}}}

\newacro{SACK}[SACK]{Selective Acknowledments}
\newacro{SMB}[SMB]{Server Message Block}
\newacro{TCP}[TCP]{Transmission Control Protocol}
\newacro{TLS}[TLS]{Transport Layer Security}
\newacro{TS}[TS]{Timestamps}
\newacro{UDP}[UDP]{User Datagram Protocol}
\newacro{WS}[WS]{Window Scale}
\newacro{XMPP}[XMPP]{Extensible Messaging and Presence Protocol}

\begin{document}

\makeatletter 
\renewcommand\Huge{\@setfontsize\Huge{17pt}{20}}
\makeatother
\title{Designing Transport-Level Encryption for Datacenter Networks}

\author{
        \rm{Tianyi Gao, Xinshu Ma, Suhas Narreddy, Eugenio Luo, Steven W. D.  Chien, and Michio Honda}\\
        \emph{University of Edinburgh}
}

\pagenumbering{arabic}

\maketitle
\thispagestyle{plain}
\pagestyle{plain}


\begin{abstract}
Cloud applications need network data encryption to isolate from other tenants
and protect their data from potential eavesdroppers in the network infrastructure.
This paper presents \sdp, a protocol design for emerging datacenter transport
protocols, such as NDP and \homa, to integrate data
encryption.
\sdp integrates TLS-based encryption with a message-based transport protocol
	that supports efficient Remote Procedure Calls (RPCs), a common workload in datacenters.
This architecture enables the use of per-message record sequence number spaces in a secure
session, while ensuring unique message identities to prevent replay attacks.
It also enables the use of existing NIC offloads designed for TLS over TCP,
while being a native transport protocol alongside TCP and UDP.
We implement \sdp in the Linux kernel by extending Homa/Linux and
improve RPC throughput by up to \SI{41}{\%} and latency by up to
	\SI{35}{\%}  in comparison to TLS/TCP.

\end{abstract}

\section{Introduction}\label{sec:intro}
Datacenter transport protocols, pioneered by DCTCP~\cite{dctcp}, have evolved over the last
decade to achieve high throughput for bulk transfer while maintaining low
latency for small messages.
The latest ones, including NDP~\cite{ndp} and Homa~\cite{homa}, are not
extensions to TCP---they are message-based and employ clean-slate
congestion control designs often with switch support to enable fine-grained
network utilization.
While most of these protocols have been implemented in user space or simulators,
the availability of the Linux kernel implementation of Homa since 2021~\cite{homalinux},
along with its demonstrated use in industry~\cite{bytedancehoma},
makes widespread adoption of alternative transport protocols within reach.

However, what if the applications want data encryption to isolate themselves from other
tenants and protect themselves from network infrastructure?
There is widespread agreement that datacenter networks need
encryption~\cite{davieeastwest}.
Major operators already adopt it, as seen in TLS deployment in
Google~\cite{enctransit} and Meta~\cite{encfb,pqfb} datacenters today.
It protects users or operators from malicious
insiders who may act as
\emph{man-in-the-middle}~\cite{mogulphysical}\footnote{The
incident indeed happened in 2013 and it accelerated adoption of
traffic encryption~\cite{gellman2013nsa}.}.
It is also essential in multi-tenant datacenters, where compromised tenant
instances may attack other tenants or the shared network
infrastructure~\cite{privateeye,netvigil}, which is often
misconfigured~\cite{mogulphysical,aura} or lacks timely security
updates~\cite{equifax}.

We present a secure message transport protocol (\sdp).
The design goal of \sdp is to achieve performance-related properties of
datacenter transports while supporting the same threat model as TLS/TCP---to
protect the endpoints from data breach, packet injection, and replay attacks.
\sdp uses per-message TLS record sequence number space in the authenticated
session, while guaranteeing message uniqueness to protect the applications from
replay attacks.
This design enables unordered encrypted messages while using existing TLS
offload and segmentation offload available in commodity NICs~\cite{autonomous}.
This means that \sdp can be adopted without compromising
hardware offload currently used by TLS/TCP.
\sdp uses plaintext message identifiers and offsets in packet headers.
This enables the network or the host stack to perform message-granularity
operations, such as load balancing across multiple paths or CPU cores.
\sdp can be a native transport protocol without relying on TCP or UDP protocol number.
This generalizes the design primitives of \sdp, allowing them to be applied to secure other
datacenter transports.

We have implemented \sdp in the Linux kernel by extending \khoma~\cite{homalinux}, because it provides a
middle ground as an unencrypted but message-based datacenter transport protocol
that could be transformed to another protocol like NDP (\autoref{sec:options}).
This paper makes two main contributions:
\begin{itemize}[leftmargin=*]
\item We identify a design point of an encrypted message-based datacenter
transport protocol that is native and compatible with existing TLS
offload while enabling the same security properties as TLS/TCP.
\item We provide a proof-of-concept implementation of \sdp that
	exhibits at most 41\% higher throughput than TLS/TCP. We also report the application
porting effort to use \sdp through two applications: Redis key value store
and NVMe-oF in-kernel storage subsystem.
\end{itemize}

\section{Design Space}\label{sec:options}
\begin{figure*}[tb]
\centering
         \begin{minipage}[b]{0.65\linewidth}
         \captionsetup{type=table}
	\scriptsize
        \renewcommand{\arraystretch}{0.8}
        \newcolumntype{H}{>{\raggedright\arraybackslash\setstretch{0.9}}p{15mm}}
        \newcolumntype{N}{>{\raggedleft\arraybackslash\setstretch{0.9}}p{12mm}}
        \newcolumntype{M}{>{\raggedleft\arraybackslash\setstretch{0.9}}p{10mm}}
        \begin{tabularx}{\linewidth}{H N N M N N X}
                & Encrypt. & Abstract. & Offload  & Protocol & Parallelism & \\
                \toprule
                \tcpcrypt\cite{tcpcrypt} & \tcpcrypt & Stream & TSO & TCP & Conn. & \\
                \midrule
                QUIC\cite{rfc9000} & QUIC-TLS & Stream & N & UDP & Conn. & \\
                \midrule
                TCPLS\cite{tcpls} & TLS & Stream & TSO & TCP & Conn. & \\
                \midrule
                TLS/TCP\cite{autonomous} & TLS & Stream & Enc.+TSO & TCP & Conn. &\\
                \toprule
                \textbf{\sdp} & TLS & Msg. & Enc.+TSO & New & Msg. & \\
                \toprule
                Homa\cite{homalinux}/NDP\cite{ndp} & - & Msg. & TSO & New & Msg.  & \\
                \midrule
                MTP\cite{mtp} & - & Msg. & N/A & New & Msg. & \\
                \midrule
                Falcon~\cite{falcon}/UET~\cite{uet} & PSP & Msg. & Full & UDP & Msg. & Custom NIC\\
                \midrule
                SRD\cite{srd} & - & Msg. & Full & N/A & Msg. & Custom NIC\\
                \midrule
                KCM\cite{kcm}/\utcp\cite{minion} & - & Msg. & TSO & TCP & Conn. & \\
                \bottomrule
        \end{tabularx}
         \caption{Key properties of encrypted or message-based
                 transport methods (discussed in \autoref{sec:options:enc} and
                 \autoref{sec:options:msg}).}\label{table:transports}
         \end{minipage}
         \hspace{2mm}
         \begin{minipage}[b]{0.32\linewidth}
                 \centering
         \includegraphics[width=0.4\linewidth]{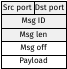}
	 \caption{Generalized message-based transport packet format
                 based on \homa~\cite{homalinux} and MTP~\cite{mtp}.
	 \normalfont{Shaded parts are identical between the packets that belong
                 to the same message. \textsf{Msg off} identifies the position
                 of this packet within the message.}} \label{fig:genhdr}
                 \vspace{2mm}
         \end{minipage}
\end{figure*}

Datacenter applications exhibit Remote Procedure Call (RPC) workloads, where
they send or receive structured messages in a request-response
manner~\cite{rpcbench,servicerouter}.
RPCs are used for a range of purposes, including API calls between
services~\cite{crisp,ubertcp}, access to in-memory key-value
caches~\cite{pegasus} or blob storage~\cite{tectonic}, and cluster management in
microservice~\cite{appnet} or FaaS ~\cite{dirigent} platforms.
RPCs are typically small (e.g., \cite{seemakhupt2023cloud} reports that half of
RPCs have median requests and responses under \SI{1530}{\byte} and
\SI{315}{\byte}, respectively) and highly concurrent.

TCP is fundamentally unsuitable for RPCs because of two reasons.
First, it disregards message boundaries that the application would have implied
with separate \texttt{send} calls.
Therefore, the application indicates the message length at the beginning of
each message, so that the receiver, which may read partial or
multiple messages at once from the bytestream, can reconstruct the original messages.
Second, in-order bytestream abstraction causes head-of-line blocking (HoLB).
It is not only triggered by packet loss or retransmission, but also on a CPU core.
Since the host network stack parallelizes ingress and egress processing
across the cores in a flow 5-tuple granularity to avoid packet reordering in a
connection, a small
message needs to wait for a preceding large one processed on the same core.

The application could increase message concurrency over parallel connections, but a
large number of connections stress both the transport layer (e.g., cache pollution with
connection metadata~\cite{megapipe,mtcp,openbypass}) and application (e.g., per-socket
syscalls~\cite{stackmap,homalinux}).
Also, parallel connections do not solve HoLB at a CPU core, once the number
of those exceeds that of cores.

Despite those shortcomings, TCP is widely used in datacenters as a convenient
reliable, congestion-controlled transport medium for RPC protocols, such as
HTTP, gRPC and Thrift, also benefiting from NIC offloading for
segmentation.
TLS encryption is also common and thus all of those RPC protocols support it.
Furthermore, container clusters often use encryption with mutual authentication (mTLS) to
interconnect services over the service mesh~\cite{cilium}.

Therefore, encrypted datacenter transport protocols must support RPC workloads
efficiently.
At the same time, it must retain the offload capabilities currently
available in TCP and TLS, which is crucial for leaving
sufficient CPU cycles for applications or improving energy efficiency, and
threat model as existing encrypted communication.
Support for those properties could facilitate departure from TLS/TCP.

In this section we explore the design space of such a transport protocol based
on those requirements in either literature or new experiments where it is
unknown.
Three key observations that guide the design of \sdp stand out:
\begin{itemize}[leftmargin=*]
\item Existing encrypted transports fail to support hardware
offload or message-based abstraction (\autoref{sec:options:enc}).
\item \homa provides a middle ground as an unencrypted message-based transport
protocol for datacenters (\autoref{sec:options:msg}).
\item TLS offload available in commodity NICs can be generalized to new transport
        protocols (\autoref{sec:options:tcpas}).
\end{itemize}

\subsection{Transport-Level Encryption}\label{sec:options:enc}

Despite some momentum to integrate encryption into a transport protocol, 
since existing approaches have been designed for the Internet
applications, they do not primarily focus on host-stack software overheads or
HoLB at a CPU core, as reviewed from the top to the middle in
\autoref{table:transports} in the rest of this subsection.

\tcpcrypt~\cite{rfc8548,tcpcrypt}, designed for the Internet before the wide adoption of
TLS, extends TCP for connection authentication and data encryption.
\tcpcrypt encrypts TCP payload with AEAD using the key exchanged during connection setup.
\tcpcrypt is unsuitable for datacenters, because it inherits the HoLB problems
in TCP and its cryptographic operations cannot be offloaded to commodity NICs.

QUIC is a transport protocol designed for the web.
It runs in userspace on top of UDP and integrates a custom version of TLS
1.3~\cite{rfc9001}.
Although QUIC mitigates HoLB on a packet loss using multiple streams in
the connection, it does not solve HoLB on a host CPU core due to
connection-level core affinity.
In addition, its complex protocol design incurs high software
overheads~\cite{quicnotfast,dcquic} and its cryptographic operations cannot be
offloaded to today's commodity NICs.

TCPLS~\cite{tcpls} provides similar features to QUIC, such as multiple streams,
but over TCP to traverse more middleboxes.
It extends the TLS 1.3 record type to aggregate and synchronize multiple TCP connections at
the TLS endpoint.
In addition to the inherent HoLB problems in TCP-based approaches (
\autoref{sec:options}), TCPLS cannot utilize TLS offload
due to its custom method of calculating the AEAD nonce~\cite{piraux-tcpls-00}.

\ktls~\cite{ktlsdoc} accelerates TLS/(DC)TCP by offloading encryption and decryption tasks to
the kernel.
It has been used by Facebook for datacenter networking~\cite{dcns}, Netflix for
 video streaming~\cite{ntflxtls} and Cisco/Cilium for network
observability~\cite{ktlscilium,ktlscilium2}.
It enables opportunistic NIC offload for cryptographic
operations, along with segmentation offload.
\ktls inherits the HoLB problems from TCP.

\subsection{Message-Based Transport}\label{sec:options:msg}

Although existing transport-level encryption approaches are unsuitable
for datacenter RPCs, there exist several attempts to enable message-based
transport abstractions mostly without encryption.
However, we must consider HoLB avoidance at both packet loss and CPU core,
high-bandwidth and low-latency datacenter networking requirements, and generality.
We review those attempts from the bottom to the middle of
\autoref{table:transports} in the rest of this subsection.

Kernel Connection Multiplexer (KCM)~\cite{kcm,dcns} provides message-based
abstractions through datagram socket APIs (e.g., \texttt{sendmsg/recvmsg}) over TCP
connections.
However, it incurs high CPU overheads for locating the framing headers in the
stream using an \ebpf program supplied by the application.
It also leaves HoLB on packet loss or CPU core unresolved.

\minion/\utcp~\cite{minion} enables TCP bytestream self-delimitation using consistent 
overhead byte stuffing (COBS) with a single delimiter byte, slightly increasing the data length.
This allows the application to retrieve out-of-order, yet meaningful data or
 messages from the kernel buffer using a new socket API.
 \utcp\@ mitigates HoLB caused by packet losses, but not the one on a CPU
 core.
It also incurs high overheads to encode or decode the data
 with COBS.

SRD~\cite{srd} is a transport implemented in a custom NIC.
While performing in-NIC multipath congestion control, it delivers out-of-order
packets to the software, which implements message abstraction.
Falcon~\cite{falcon} and UET~\cite{uet} are also hardware-based transports, but
unlike SRD, they implement message abstraction, such as message-level
reliability, in hardware.
Those transports mainly focus on GPU-based HPC/AI workloads.
For wider deployment scenarios, we seek an approach that is compatible with
commodity NICs and can be used in bare metal or virtualized cloud instances and
networks that currently use TLS/TCP.

\homa~\cite{homalinux,homa} is a receiver-driven transport protocol that
preserves message boundaries and mitigates HoLB on packet losses via
out-of-order message delivery. 
It also mitigates HoLB on a CPU core using shortest remaining processing time
(SRPT) scheduling, dynamically distributing messages across cores within the
same flow 5-tuple instead of binding them to a fixed core.

\autoref{fig:genhdr} depicts the simplified packet format of \homa; it also
applies to MTP~\cite{mtp}, another message-based transport designed for
in-network compute (\autoref{sec:discuss}).
To support arbitrary-sized, unordered messages, each packet contains message ID,
message length and message offset, so that the receiver can reassemble the
 messages.
Although \homa uses a new protocol number, its packet \emph{overlays}
a TCP header to utilize TCP Segmentation Offload (TSO), where a NIC splits a
large segment (called TSO segment) into MTU-sized packets.
\homa embeds the message ID in the TCP options space, which is copied to all the
packets by TSO.
It prepends the message offset to each packet payload, which is possible because
the boundaries of packets generated by TSO are predictable.
This is also necessary, as TSO does not write sequence numbers for undefined
transport protocols~\cite{homa,eqds}.

We believe \homa is a practical basis for a message-based transport protocol for datacenters 
in terms of abstraction and packet format.
\homa's host stack could be adapted to other message-based transports.
For instance, NDP~\cite{ndp} shares similar stack and protocol requirements, such as
packet scheduling for prioritizing specific data/control messages and
first-RTT data transfer. NDP packet types map naturally to those of \homa:
\textsf{NACK} in NDP and \textsf{RESEND} in \homa both request retransmission, while
their \textsf{PULL} and \textsf{GRANT} request the next data.
Also, \homa is well documented and in active development for Linux
upstreaming process~\cite{homaup}.

\subsection{Encryption Method}\label{sec:options:tcpas}

The choice of encryption method is crucial for designing a viable encrypted
datacenter transport protocol.
Of particular relevance is the deployment model and hardware offload.

IPSec provides host-to-host or site-to-site security as it operates at the
network layer configured by the operator rather than the applications.
This model thus differs from TLS whose authenticated sessions are established between
individual applications.
PSP~\cite{psp}, a more recent proposal for dataceneters, also performs packet-based
encryption but in a more scalable way than IPSec, offering
connection-granularity security.
However, PSP needs specific NICs and it does not assume software-based
encryption, because TLS is faster in software~\cite{psplinux}.
We wish to support various deployment models, much like the current TLS/TCP
ones where TLS encryption can be decided by the application and its
cryptographic operations can be optionally offloaded to the NIC when the NIC is
trusted.

Furthermore, IPSec and PSP approaches are incompatible with confidential computing
executed inside Trusted Execution Environment (TEE).
TLS is used in such use cases~\cite{fastconfidential,rktio,scone}.
Transport-level integration could be compatible with them, so long as
the protocol is implementable in user-space.

Therefore, TLS appears the best option for smooth transition from TLS/TCP.
However, a big question is whether it can be used for new, non-TCP transport
protocols, such as \homa.
We do not take this question just because \homa is a native transport protocol,
but we believe enabling encrypted datacenter transport as a native transport
makes emerging protocol design and deployment flexible.
Although attempts have been made to repurpose the TCP protocol number for a
non-TCP protocol to use NIC TSO (e.g., STT~\cite{davie-stt-05}), this approach
would not gain widespread acceptance, as it 
complicates operation of network management or monitoring
systems~\cite{privateeye} and port number management in the host TCP
implementation.

The key aspect of assessing the feasibility of using TLS with a new transport
protocol is whether TLS offload in commodity NICs can be used, because hardware
offloading is crucial for leaving as many CPU cycles as possible for
applications or improving energy efficiency.
We believe middleboxes in datacenters (e.g., load balancers) are more
evolvable than those in the Internet, because they are made of software
developed by the operator~\cite{maglev,snap} or service provider closely working
with cloud operators, whereas home gateways~\cite{hatonen2010experimental}
and appliances in access networks~\cite{exttcp,edeline2019bottom} are hard to
enforce upgrade.
NIC offload is also crucial to facilitate transition from TLS/TCP.
If the operators or applications have to give up the hardware offload currently
used for TLS/TCP, it would make the new transport slower than TLS/TCP
accelerated by the offload.
This would create a catch-22 situation, motivating no hardware vendor to support
the acceleration of the new transport protocol.

We review two existing cryptographic NIC offload architectures with
experimental validation.

Chelsio T6 released in 2016 supports TLS offload but strips TCP
options provided by the stack, as it relies on the TCP full offload engine (TOE).
It is thus unsuitable for not only new transport protocols but TCP extensions, a limitation 
noted by Netflix, Microsoft, and others~\cite{autonomous}.

In contrast, NVIDIA ConnectX-6 DX (CX6) and -7 (CX7), released in
2020 and 2023, respectively, feature a different hardware architecture, called 
\emph{autonomous offload}~\cite{autonomous}.
This architecture allows the transport protocol to run in
software, allowing it to evolve, while offloading data processing in an application-level protocol like TLS.
Linux has supported this architecture, and its software interfaces and hardware requirements for other
 vendors are documented~\cite{ktlsdoc}.
These NICs are widely used today, with NVIDIA holding the largest NIC market
share for NICs supporting \SI{25}{\Gbps} and above (e.g., \SI{65}{\%} in 2019~\cite{nvidiashare}).
The distinctive software interfaces described in \cite{ktlsdoc} allow us to 
infer the TLS offload
 architecture of other NICs in their Linux drivers.
Broadcom, Microsoft/Fungible and Netronome NICs appear to support this architecture,
while Intel might not.

We tested CX6 and CX7 NICs by generating a TLS/TCP TSO segment using \ktls.
In the driver, we modified the protocol number field in the IP header just
before the packet descriptor was linked to the hardware.
We confirmed that the resulting packets have correctly encrypted payload while
preserving the original TCP header structure with or without TSO.
This observation indicates feasibility of enabling a new encrypted transport protocol that 
can benefit from \emph{existing} hardware acceleration.

\section{\sdp Design Challenges}\label{sec:challenges}

\sdp focuses on message-based socket abstractions where the application sends
multiple independent messages in parallel and the receiver can process them in
any order, while ensuring reliable message delivery through packet retransmissions.
\sdp provides TLS-based security guarantees for such an abstraction implemented
by Homa~\cite{homalinux}, which achieves datacenter-friendly properties of 
RPC efficiency, host stack parallelism, and generality to extend to other
message-based datacenter transports (\autoref{sec:options:msg}).

Achieving those datacenter transport properties while adding security is
challenging due to the TLS protocol semantics and stack and NIC features.

\subsection{TLS Protocol Semantics}\label{sec:challenges:tls}

TLS assumes in-order bytestream abstraction for the underlying transport and
guarantees the original order of the records, rejecting out-of-order or duplicated records, which would
have been tampered or replayed but delivered by TCP due to TCP-level correctness
based on sequence number and checksum.
This means that simply \emph{stacking} TLS over a message-based transport like
\homa is not viable, because out-of-order message delivery to the TLS endpoint
causes record rejection, whereas performing a TLS handshake for every message is
 impractical.

Stacking TLS over a message-based transport also precludes TSO performed
together with TLS offload.
Enabling message-based abstractions with TSO requires that the transport layer
place framing headers in the middle of the message (\autoref{sec:options:msg}),
whereas the NIC TLS offload cannot exclude such ``gaps'' from encryption.
If those framing headers were encrypted, the transport protocol could not
reassemble the TLS records from packets.

\subsection{Host Stack and NIC Features}\label{sec:challenges:host}
Message-based transport (\autoref{sec:options:msg}) could send multiple
independent messages in any order by the scheduler or
congestion control algorithm within the same flow 5 tuple.
This is a stark contrast to TCP, which serializes all the transmissions,
including retransmissions, to minimize packet reordering.
TCP transmits packets in the syscall (e.g., when a new data is written
by the application and the window is available) or interrupt
(softirq) context (e.g., when a received ack packet triggers transmission of new
data in the send buffer), both of which are performed while locking the socket.

However, message-based transport would take message-level locking without
socket-level one for message-level parallelism within the stack, as done in \homa.
Further, receiver-driven transport protocols, such as NDP and \homa, run a
dedicated packet scheduler thread for fine-grained network utilization.
For example, \homa sends small messages directly in the syscall context, but parts of
large messages are pushed by the scheduler.
When the \homa sender receives a \emph{Grant} packet, in which the receiver
grants the sender transmission of new data, it sends data chunks in the softirq context.

Those stack features pose challenges in using TLS offload in the NIC.
\begin{figure}[tb]
\centering
	\includegraphics[width=\columnwidth]{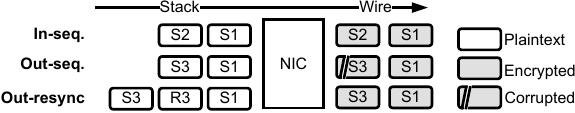}
	\caption{Encryption with autonomous
	offload~\cite{autonomous}. \normalfont{Each rectangle represents one TLS record that
	contains one or more packets or TSO segments. The
	HW expects \textsf{S2} after \textsf{S1} to
	produce a correct \emph{next} encrypted segment (\textsf{In-seq}); if
	\textsf{S3} arrives,
	it generates a corrupted one (\textsf{Out-seq.}). A resync
	descriptor (\textsf{R3}) changes the seqno that HW expects to
	\textsf{S3} (\textsf{Out-resync}). Note that each segment in the \textsf{Wire}
	actually consists of multiple packets split by TSO.}}
	\label{fig:ao}
\end{figure}
Autonomous Offload (AO) (\autoref{sec:options:tcpas}) maintains a flow context
backed by in-NIC memory, which stores the encryption key and self-incrementing
record sequence number.
\autoref{fig:ao} illustrates how AO works.
When the software sends a segment that the NIC needs to encrypt with a
different record sequence number than its current internal one, it must
prepend a \emph{resync} descriptor in the queue (\autoref{fig:ao} bottom) to
adjust that internal one.
TCP uses this feature for retransmissions where the NIC sees the
previous record sequence numbers.

Message-based transports could send multiple messages across different CPU
cores, which push their packets to different NIC queues.
This makes enforcing the NIC to encrypt a record with a specific or predictable
record sequence number hard.
Prepending a resync descriptor to every segment (to reset the sequence number
expectation from another message) does not solve the problem, because the NIC
provides no atomicity or ordering guarantee in reading the descriptors across the
queues.
Consider two segments that belong to different messages but to the same 5
tuple, \textsf{S4} and \textsf{S5} (not illustrated).
They are sent in parallel by different CPU cores (e.g., scheduler and softirq)
and thus to different NIC queues.
Although each segment prepends a resync descriptor (see
\autoref{fig:ao}), \textsf{R4} or \textsf{R5}, it is not guaranteed that
the descriptor pair of resync and segment (e.g., \textsf{R4} and
\textsf{S4}) are read by the NIC atomically;
the NIC could read \textsf{R4} after \textsf{R5} then read
\textsf{S5}, resulting in incorrect encryption.

\section{\sdp Design}\label{sec:design}

\sdp addresses the aforementioned challenges by transport-level encryption,
where the transport protocol \emph{embraces} encryption based on TLS.
This architecture enables two key features of \sdp: message format that can use both TSO and TLS offload (\autoref{sec:design:enc}) and the use of per-message
record sequence number space in the secure session for unordered message
delivery without costly per-message handshake (\autoref{sec:design:para}).
We provide detailed security analysis in \autoref{sec:sec}.

\subsection{Threat Model}\label{sec:design:threat}
We assume the same threat model as TLS/TCP, protecting endpoints
from data breaches, packet injection, and replay attacks.
We assume the host subsystem that executes the transport protocol---
the OS kernel in our implementation---is trusted.
When the OS kernel cannot be trusted, \sdp can be implemented in user-space
protected by a TEE environment---using a trusted network stack like \rktio~\cite{rktio}.
While we also assume the NIC is trusted, this assumption can be
removed; in such cases, TLS offload must be disabled so that the NIC processes
only encrypted packets.

\subsection{Session Initiation}\label{sec:design:session}
\sdp initiates a secure session using the standard TLS 1.3 handshake
performed by the application, because datacenter transport protocols, such as
Homa and NDP, send an RPC already on the first RTT without transport-level
handshake.
A session is identified by the flow 5 tuple that consists of source-destination
address and ports plus protocol.
Since the handshake process is based on TLS 1.3, it can support mutual
authentication as with mTLS~\cite{ciliummtls}.

After the handshake, the application registers the initialization vectors and
session keys negotiated over the handshake to the \sdp socket\footnote{Same as \ktls:
\url{https://docs.kernel.org/networking/tls.html}.}.
After that, a plaintext message written to the socket is encrypted and sent by \sdp.
The \sdp receiver decrypts the message and the application reads the plaintext one.

Although the session initiation takes one RTT, the application can reuse the
same shared key over multiple, concurrent messages in the session for a while.
Alternatively, it can also be done over the first RTT \sdp data at the expense
of forward secrecy of that data (not subsequent ones), which we discuss in \autoref{sec:design:key}.

\subsection{Offload-Friendly Encrypted Message Format}\label{sec:design:enc}
\begin{figure}[tb]
\centering
	\includegraphics[width=0.75\linewidth]{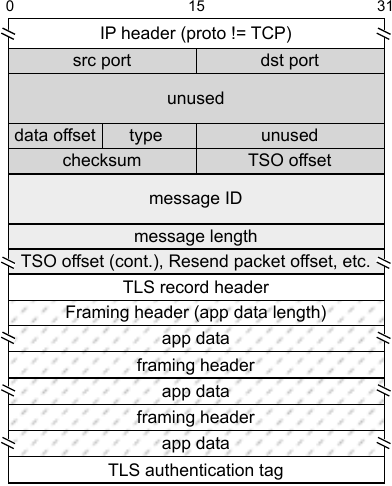}
	\caption{\sdp TSO segment with one TLS record being split to 3 packets.
	\normalfont{Dark and
	light gray parts overlay TCP common header and options space,
	respectively, and are replicated over every packet by TSO. The NIC
	encrypts the dashed area. TLS record header is actually \SI{5}{\byte}
	and the authentication tag is \SI{16}{\byte}.}}
\label{fig:sdpgso}
\end{figure}

\sdp uses the TCP header structure with a new transport
protocol number indicated in the network layer header to use TSO, like
\homa~\cite{homalinux}; we confirmed that it can also be compatible with TLS
offload in \autoref{sec:options:tcpas}.
When the application posts a message to an \sdp socket, the question is how to
segment this message, which can span across multiple TSO
segments or TLS records, into packets while applying encryption, possibly by the NIC.
It must be done such that the receiver can reassemble the original message from
them.
This is not a concern for TLS over TCP, because, when a TLS receiver sees a series of records,
they are already in-order based on the underlying bytestream abstraction.
However, message-based transports with TSO and TLS offload place unusual demands,
as discussed next.

\sdp segments an application message in two stages---to TSO segments
(\autoref{sec:options:msg}) and packets.
The receiver reassembles the message in the reverse order.
\sdp creates TLS records, each of which is preceded by a record header and is at
most \SI{16}{\KB} in size, to align with the boundaries of the TSO segments,
which are at most \SI{65}{\KB}.
When the NIC does not support TLS offload, the records are encrypted by the CPU at this
stage.

Each TSO segment has \emph{TSO offset}, which indicates its position within the
message.
Since \sdp embeds it in the overlayed TCP header, which is copied to all the
packets by TSO, all the packets that have been generated from the same TSO
segment have the same TSO offset value.
\sdp needs \emph{packet offsets} for the receiver to reassemble the TSO segment from the
packets.
We use the IPID in the network header because it is incremented over the packets
generated by TSO.
If the NIC generates TCP sequence numbers for non-TCP packets when performing
TSO, we could use that, as it also works for IPv6.

As a simple example, \autoref{fig:sdpgso} illustrates a message that
consists of one TSO segment and TLS record splits into three packets.

The receiver first reassembles a TSO segment based on the packet offsets in the
tuple of TSO offset and message ID.
It then decrypts the TLS records and reassembles the message based on the TSO
offset values.

We must handle two cases of retransmissions: actual packet loss and spurious retransmission, which
must be ignored by the receiver.
Retransmission of a packet needs to have the original packet offset within the segment,
we embed that value in the unused space of the overlayed TCP header
(\textsf{Resend packet offset} in \autoref{fig:sdpgso}, plaintext area).

Note that the use of framing headers is based on our current implementation.
We could remove it, because the receiver can reassemble the TSO segments using
solely packet offsets.
This would improve performance of large messages because of simpler buffer
operations.

\subsection{Per-Message Record Sequence Number Space}\label{sec:design:para}
To avoid costly handshake performed for every message or record rejection caused
by out-of-order message delivery (\autoref{sec:challenges:tls}), \sdp uses
per-message record sequence number space within the TLS session.
Each record sequence number space offers the order-preserving guarantee of
TLS on top of reliable message delivery of message-based transports like \homa
and NDP.
The record sequence number monotonically increments in the message like regular TLS.
Record sequence number spaces are mapped to the message IDs.
When the receiver sees the first packet that belongs to an unseen
message ID, it initializes the next in-sequence record sequence number.

\begin{figure*}[tb]
  \centering
  \includegraphics[width=\textwidth]{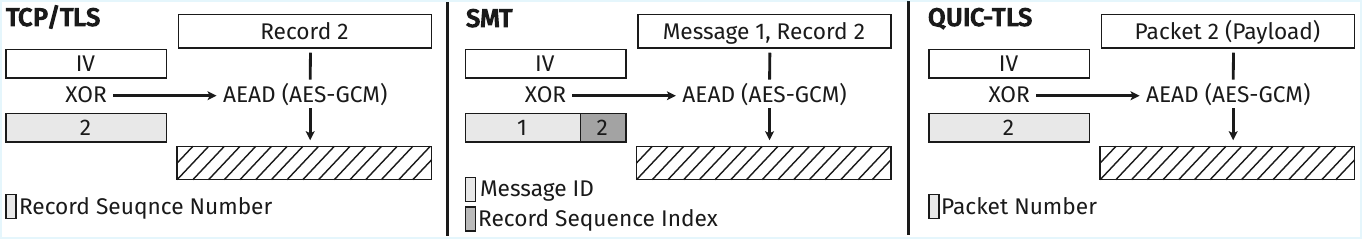}
  \caption{Use of record sequence numbers across TCP/TLS, \sdp, and
  QUIC-TLS~\cite{rfc9001}:
  TCP/TLS uses the 64-bit record sequence number;
  \sdp encodes message ID and intra-message record index
  (\autoref{sec:design:msgid});
  QUIC uses the packet number (\autoref{sec:other}).}
  \label{fig:rec_seq_nounce}
\end{figure*}

\begin{figure}[tb]
  \centering
  \includegraphics[width=\columnwidth]{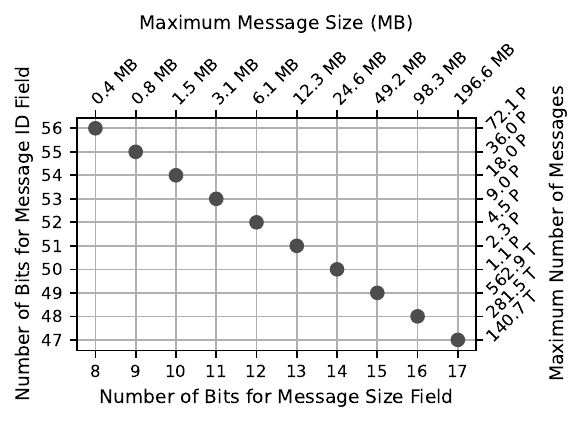}
      \caption{Trade-off between maximum message size and number of unique
      message IDs based on bit allocation in the 64-bit
      composite record sequence number.}
      \label{fig:bit_allocation_tradeoff} 
\end{figure}

However, this parallelism introduces challenges for TLS which is
designed for TCP abstraction.
TLS requires a unique record sequence number for each record in one
handshake to prevent replay attacks.
However, using multiple (i.e., per-message) record sequence number spaces itself means that the receiver may see the
same record sequence number between the messages.

\subsubsection{Message Uniqueness Guarantee}\label{sec:design:msgid}
Transport-level encryption, instead of stacking TLS on a non-encrypted transport
protocol, enables solving this issue.
\sdp introduces a composite 64-bit record sequence number that
 integrates a message ID whose uniqueness is guaranteed throughout the secure session
 with an intra-message record index.

To comply with TLS, TLS record sequence number with fixed 64 bits length is the
only free variable available to encode both message ID and intra-message record index.
We dedicate a portion of these bits to message IDs and assign the remaining bits
to indexes of the records within the message (intra-message record index) as
illustrated in \autoref{fig:rec_seq_nounce}, It requires bit allocation trade-off
between maximum message sizes and the number of unique message IDs.
Since the maximum record size is \SI{16}{\KB}, supporting larger maximum message
sizes needs more bits for the record indexes; supporting more messages needs
more bits for the message IDs.
\autoref{fig:bit_allocation_tradeoff} plots this trade-off.

In our current implementation, we opt for 48-bit message IDs.
This leaves 16 bits for the intra-message record index.
This allocation allows a single message to accommodate up to 65K individual TLS
records, supporting message sizes up to approximately \SI{98}{\MB} even with
\SI{1.5}{\KB} (small) TLS records, and approximately \SI{1}{\GB} with
\SI{16}{\KB} one (maximum record size).
For reference, the default maximum message size of \homa is \SI{1}{\MB}.
Each endpoint can use different message ID length as long as the receiver
endpoint knows what the sender uses. That could be negotiated during the
handshake.
Revealing the message ID length to an eavesdropper does not increase the security
risk.

The composition itself introduces very little performance
overhead, because the intra-message record index occupies the lower bits, allowing
the hardware's self-incrementing counter to operate correctly
just like TLS/TCP.

\subsubsection{TLS Hardware Offload}\label{sec:design:offload}

Per-message record sequence number spaces enable the use of AO-based TLS
offload, avoiding the problem with non-atomic reads between the descriptors
across the queues (\autoref{sec:challenges:host}), because
the messages in the same 5 tuple do not have to share the flow context, which 
dictates in-sequence record sequence numbers, across the queues.
This approach also enables efficient use of in-NIC memory, because it allows a flow context
to be reused by another message in the same session (i.e., record sequence
number space) simply performing a resync operation.
This is not the case when switching the keys (e.g., with another handshake); it
requires allocation of a new flow context, which is more expensive than
resyncing existing one.

Although the NIC can maintain millions of active flow contexts in its memory and
packet transmissions effectively hide the cache evictions and
admissions~\cite{autonomous,nicmem}, it cannot maintain an unlimited number of
them.
Therefore, taking the advantage of efficient reuse of the context, we design a
flexible trade-off between parallelism and in-NIC memory usage.
We create flow contexts for each message until a certain threshold, at least
one per NIC queue.
Messages that go to the same queue in the same flow 5 tuple may share the same
context, but those sent to different queues do not.
However, even when sharing the context, since they are serialized in the
queue, a resync operation guarantees its target segment or record.
Note that the segments that constitute the same message always go to the same queue, because
the message-based transport could ensure in-order delivery \emph{within} the
message to avoid packet-level reordering.

Our current implementation allocates one flow context per queue for each flow 5
tuple, but we may revise this in the future or for other NICs.

\subsection{Key Exchange and 0-RTT Data}\label{sec:design:key}
Efficient key exchange is essential for cloud applications to initiate data
transfer with minimal latency. While TLS 1.3, the default key exchange method
for \sdp, supports fast resumption and \sdp can send multiple messages in
the same session (\autoref{sec:design:session}), its effectiveness
decreases in dynamic communication patterns where endpoint churn limits session
reuse.
To understand the overhead of key exchange, we
take the latency breakdown of the TLS 1.3 initial handshake by timestamping the \texttt{picotls}
library (\autoref{table:tlsbreakdown}). To accelerate
key exchange, we first introduce three techniques that help reduce handshake
latency (\autoref{sec:design:faster}). We then show how to eliminate 1-RTT by
pre-distributing long-term public key shares to internal DNS resolvers in
the datacenter (\autoref{sec:design:key:rtt}).

\subsubsection{Key Management and Authentication}\label{sec:design:faster}
\noindent\textbf{\\Key pre-generation.}
To reduce costs in S2.1 and C1.1, servers and clients could maintain a list of standby
key pairs created prior to a handshake~\cite{eng25519}.
This is feasible in datacenters that centralize administrative control such that
a choice of security parameters is made upfront.

\bh{ECDSA authentication}
ECDSA significantly reduces handshake latency—by hundreds of $\mu$s in S2.5, C4.1,
 and C4.2—particularly in mutual authentication, where the server performs one
 Sign and two Verify operations.

\bh{Short certificate chain}
To reduce latency in C3.2, we could use a short certificate chain and configure all
endpoints with the CA’s verification key, avoiding certificate lookup and
long-chain validation. This speeds up the Verify Cert operation by approximately
\SI{52}{\%} in our tests.
Since an internal CA manages certificates within the datacenter,
backward compatibility features can also be omitted.

\begin{table}
  \footnotesize
  \renewcommand{\arraystretch}{1.2}
  \newcolumntype{P}[1]{>{\centering\arraybackslash}p{#1}}
  \begin{tabularx}{\columnwidth}{P{2.0cm}p{0.4cm}p{2.4cm}c}
    \multicolumn{1}{c}{\textbf{Server}}&\multicolumn{1}{c}{ID}&\multicolumn{1}{c}{Operation}&\multicolumn{1}{c}{Overhead (\us)}\\
    \toprule
    Handle CHLO &S1 & Process CHLO& 1.8\\
    \hline
    \multirow{4}{*}{Generate SHLO} & S2.1 & Key Gen& 67.9\\
    &S2.2 & ECDH Exchange & 265.0\\
    &S2.3 & SHLO Gen & 75.2\\
    &S2.4 & EE \& Cert Encode & 13.6\\
    &S2.5 & CertVerify Gen & 137.6$^*$ / 1344.0$^+$\\
    &S2.6 & Secret Derive &48.6\\
    \hline
    Handle Finished &S3 & Process Finished & 44.4\\
\end{tabularx}
    \begin{tabularx}{\columnwidth}{P{2.0cm}p{0.4cm}p{2.4cm}c}
    \multicolumn{1}{c}{\textbf{Client}} & \multicolumn{1}{c}{} & \multicolumn{1}{c}{} & \multicolumn{1}{c}{}\\
    \toprule
    \multirow{2}{*}{Generate CHLO} &C1.1 & Key Gen& 61.3\\
    &C1.2 & Others Gen& 5.5\\
    \hline
    \multirow{2}{*}{Handle SHLO} &C2.1 & Process SHLO& 2.6\\
    &C2.2 & ECDH Exchange& 88.7\\
    &C2.3 & Secret Derive& 48.8\\
    \hline
    \multirow{2}{*}{Verify Cert}&C3.1& Decode Cert& 0.1\\
    &C3.2 &Verify Cert\tnote{1}& 483.4\\
    \hline
    \multirow{2}{*}{Verify CertVerify}&C4.1 &Build Sign Data & 1.4\\
    &C4.2 & Verify CertVerify& 196.3$^*$ / 67.1$^+$\\
    \hline
    Handle Finished &C5 & Process Finished & 42.6\\
\end{tabularx}
  \caption{Server- and client-side TLS handshake overheads ($*$ with 256-bit
	ECDSA and $+$ with 2048-bit RSA).}
  \label{table:tlsbreakdown}
\end{table}

\subsubsection{0-RTT Data and Key Exchange}
\label{sec:design:key:rtt}
 Datacenter transport protocols, such as Homa and NDP, send an RPC already on the
 first RTT without transport-level handshake.
0-RTT data could be achieved by extending the TLS 1.3 key exchange with DNS-based distribution of a server’s long-term Diffie-Hellman (DH) public key. This approach, inspired by TLS Encrypted Client Hello (ECH)~\cite{ietf-tls-esni-17}, removes one RTT from the initial handshake.

In our design, the client first performs a DNS query to retrieve
\textit{\sdp-ticket}, which includes: (\romannum{1}) the server’s long-term ECDH
public key share, (\romannum{2}) its certificate, and (\romannum{3}) a
signature over the \textit{\sdp-ticket}s signed by the certificate’s private key.  Note that the datacenter or cloud provider could operate its own root CA that also acts as the internal DNS resolver. With trusted CA public key pre-installed across the datacenter, the client can verify the \textit{\sdp-ticket} and send a ClientHello with its ephemeral key. These steps can occur before the handshake begins, as server information is often known in advance.

Using the server’s long-term key and its own ephemeral key, the client derives
an \emph{\sdp-key} and immediately sends encrypted application data.
If forward secrecy is enabled, the server replies with a ServerHello containing its
ephemeral key, enabling both sides to derive an \emph{fs-key} and switch to
forward-secret encryption. If forward secrecy is disabled, the \emph{\sdp-key}
is used for all payload encryption for the duration of the session.

We retain TLS 1.3’s session resumption mechanism, which updates cryptographic
keys and thus resets the message ID space.

\subsubsection{Forward Secrecy}

The 0-RTT handshake trades some forward secrecy for lower latency, as client 0-RTT
data is encrypted using the \emph{\sdp-key}, which lacks strong forward secrecy.
To mitigate the risk, we limit \textit{\sdp-ticket} validity period.
Following industry practice—such as Cloudflare’s hourly rotation of session ticket keys
for 0-RTT data~\cite{cloudflaretls}—we recommend a maximum lifetime of one hour.
To further reduce replay risk caused by \emph{\sdp-key}, servers can record the CHLO random value, as
specified in TLS 1.3~\cite{rfc8446}.

\subsection{Implementation}
The current \sdp implementation\footnote{\url{https://github.com/uoenoplab/smt}.} consists of a 2800 LoC patch to the Homa/Linux
kernel module and a 300 LoC patch to the NVIDIA \texttt{mlx5} driver,
requiring only these two kernel modules to be recompiled and reloaded.

Note that the device driver modification is to adjust the offset to start the
encryption specifically for TLS offload and generalize flow context management
that currently relies on TCP sequence number and lacks sufficient flexibility
of allocation.
We therefore believe its adoption once \homa is upstreamed~\cite{homaup}.

To implement the key exchange method in \autoref{sec:design:key}, we extend
\texttt{picotls} with a new extension, to indicate the use of \sdp-ticket,
reusing the \textsf{pre\_shared\_key} field to specify its identity in the
handshake.

\section{Evaluation}\label{sec:eval}
We measure the performance of \sdp in comparison to TLS/TCP and other systems.

\bh{HW\&OS}
We use two identical machines connected back-to-back.
Each machine is equipped with two Intel Xeon Silver 4314 CPUs and
NVIDIA/Mellanox ConnectX-7 \SI{100}{\Gbps} NIC.
They install Linux kernel 6.2.
We use one NUMA node, and separate cores for softirq contexts and application
threads.
The network MTU size is \SI{1.5}{\KB} unless otherwise stated.
All experiments use AES-128-GCM (128-bit length key) cipher for both \sdp and TLS.
We don’t use receive-side offload for kTLS, because not only \sdp does not
support it (\autoref{sec:discuss}), but it often impractical due to
incompatibility with tunnelling protocols or packet delivery delay when the NIC
waits for the complete record.

\bh{Performance metric}
Our primary performance metric is the protocol and encryption overhead added to the
base unencrypted variant (i.e., \homa), which we compare with that of TLS over TCP.
This makes our measurements worthwhile, although \homa variants often
do not perform better than the TCP variants due to immaturity of \homa itself,
because those characteristics could still be valid even after the \homa implementation
evolves.
Later in this section, we also provide a snapshot of the performance of \homa
and TCP variants with other aspects of \sdp evaluation, which include application
porting effort.

\subsection{Unloaded RTT}\label{sec:eval:unloaded}
\begin{figure}[tb]
\centering
	\includegraphics[width=\columnwidth]{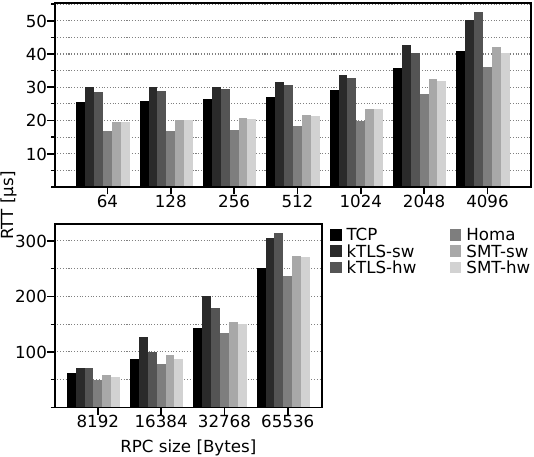}
        \caption{Unloaded RTTs of various sized RPCs. Standard deviations are
        \SIrange{2}{12}{\%}.}
\label{fig:unloaded}
\end{figure}

\begin{figure*}[tb]
\centering
	\includegraphics[width=\linewidth]{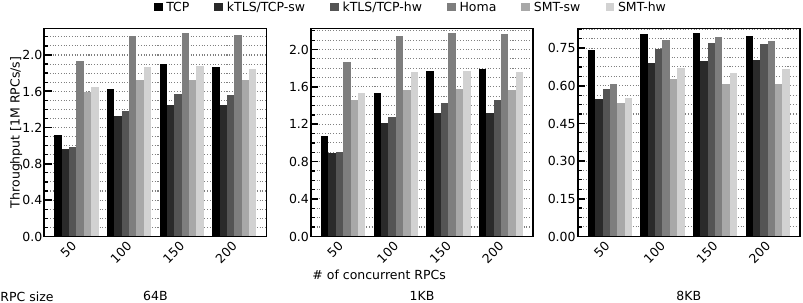}
	\caption{Concurrent RPC throughput.}
\label{fig:multi}
\end{figure*}

We first measure RTT of a single RPC without concurrent RPCs, using our custom
application to highlight software overheads of the network stack, including
the transport protocol, without the effect of queuing or application-level
processing delays.
\autoref{fig:unloaded} shows the results.
We ran three trials, 8 seconds each, and plot the middle one in average latency;
same for the next experiment.

\sdp outperforms kTLS by \SIrange{13}{32}{\%} with TLS offload and
\SIrange{10}{35}{\%} without it.
Since Homa is faster than TCP by \SIrange{5}{35}{\%}, \sdp does not diminish the
advantage of \homa over TCP.
The margin is smallest with \SI{65}{\KB} RPCs, because the \homa receiver waits for the
 arrival of the entire RPC (that consists of multiple packets) before
 delivering (copying) data to the application, whereas TCP overlaps packet
 reception and application-data delivery due to its streaming abstraction.
 This is not a fundamental limitation of \homa and there exists work to remedy
 this issue~\cite{bytedancehoma,homarecvbuf}, and once it is merged into \homa,
 we expect that \homa and \sdp outperform TCP variants by larger margins.

The benefit of hardware offloading is small (up to \SI{7}{\%} in \sdp) in this
experiment.
For small messages, this is due to low encryption overheads and per-segment
cost incurred to populate offloading metadata
(\autoref{sec:challenges}).
For large messages, the bottleneck is not encryption but data
copy.
To confirm the similar characteristics of larger message, we experimented with
\SI{500}{\KB} RPCs (not plotted); it exhibited little (\SI{1}{\%}) latency
benefit of hardware offloading.
In the next experiments where CPU cores are more loaded due to concurrent RPCs
(\autoref{sec:eval:concurrent}) or complex application-level processing
(\autoref{sec:eval:redis}), we observe a larger benefit of hardware offloading.

\subsection{Throughput}\label{sec:eval:concurrent}
Next, we measure the performance of \sdp in the presence of concurrent RPCs and multiple application threads.
We generate concurrent RPCs using 12 threads allocated for the applications and
4 threads for the stack at each of the server and client.
Recent report~\cite{seemakhupt2023cloud} shows \SI{90}{\%} of RPCs in production
are smaller than \SI{10}{\KB}. We thus evaluate three representative
sub-\SI{10}{\KB} sizes---small, near-MTU, and multi-MTUs.

\autoref{fig:multi} shows the throughput over different numbers of concurrent
RPCs for three RPC sizes.
For \SI{64}{B} messages, \sdp exhibits higher throughput than \ktls by
\SIrange{16}{40}{\%} with TLS offload and \SIrange{16}{40}{\%} without it; those
improvements with \SI{1}{\KB} messages are \SIrange{17}{41}{\%} and
\SIrange{16}{39}{\%}, respectively.

\sdp exhibits lower throughput than \ktls with \SI{8}{\KB} messages, by
\SIrange{5}{15}{\%} with TLS offload and \SIrange{3}{13}{\%} without it,
because, as before, \homa is unoptimized yet for large messages in comparison to
TCP.

In \sdp, advantage of HW is largest with 1KB cases (\SIrange{5}{11}{\%}),
because \SI{8}{\KB} cases (\SIrange{4}{9}{\%} improvement), throughput are
constrained by the lack of pipelining.

In \SI{64}{B} RPCs with 50--100 concurrent requests, \sdp exhibits a slightly
larger benefit of HW than kTLS/TCP due to lower protocol overheads of the base
transport (\homa) than TCP (as seen in \autoref{sec:eval:unloaded}) and thus
higher relative crypto overheads.

\bh{Impact of a larger MTU}
We ran the same tests as \autoref{fig:multi} right (50–150 concurrent 8KB RPCs)
with \SI{9}{\KB} MTU (thus one message fits into a single packet).
Compared to \SI{1.5}{\KB} MTU cases, \sdp exhibited 13–28\% and
16–31\% higher throughput with and without TLS offload, respectively,
because of the reduced number of packets per message.

\bh{CPU usage}
We tested CPU usage with the setting of \autoref{fig:multi} middle, but limiting
the RPC rate of all the systems to 1.2Mreq/s to measure the resource usage over
the same request rate. \sdp-SW exhibited \SI{3.5}{\%} lower CPU usage than
kTLS-SW at the client and \SI{10.5}{\%} at the server. \sdp-HW exhibited
\SI{2}{\%} lower CPU usage than kTLS-HW at the client and \SI{8}{\%} at the server.
\sdp-HW reduced the CPU usage of \sdp-SW by \SI{4}{\%} at the server and
\SI{1.5}{\%} at the client.
Our current implementation does not show memory saving with hardware offload,
because its software encryption is done in-place.

\subsection{Redis}\label{sec:eval:redis}
\begin{figure*}[t]
\centering
	\includegraphics[width=\linewidth]{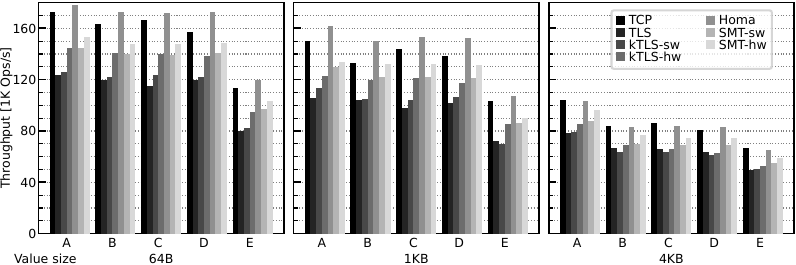}
        \caption{Redis throughput on YCSB A (update-heavy), B (read-mostly), C
        (read-only) and D (read-latest) workloads.}
\label{fig:redis}
\end{figure*}
What does using \sdp in a real-world application look like?
We report our experience of adding support for \sdp in Redis, a widely used key-value store.
The vast majority of effort was supporting vanilla \homa; once it is done,
support for \sdp was trivial because of transport-level encryption; it simply adds \texttt{setsockopt} to register the key (\autoref{sec:design}).
We thus mainly discuss adding support for \homa in Redis.

Redis adopts a single-threaded design and monitors clients using an
\texttt{epoll} event loop; each client connects to the server over a TCP
connection and the connection is reused over multiple requests.
Since a \homa socket can communicate with multiple clients, Redis/\homa could
directly block on \texttt{recvmsg} syscall.
However, to share the same database between both TCP and \homa
clients, we register the \sdp socket (that handles all the \sdp clients) to the
\texttt{epoll} socket.
When a request arrives over TCP, the Redis server reassembles messages by locating the Redis headers in the
bytestream; when that arrives over \homa, since \homa preserves message boundaries, Redis/\homa does
not need to maintain the partial read offset.

Our modification in Redis is straightforward, because \homa and \sdp
provide true file descriptors and the Redis instance can monitor both TCP and
\sdp clients in the original \texttt{epoll} event loop.
This is not the case for supporting a kernel-bypass TCP stack as done in
mTCP~\cite{acceltcp}, Demikernel~\cite{demikernel} and Paste~\cite{paste}, which
 need to replace the whole event loop, disallowing the
clients to access the same database over the regular kernel TCP stack.
The same goes for TCPLS; its I/O descriptors
cannot be registered to the OS-driven event loop.

\bh{Results} \autoref{fig:redis} shows throughput measured by
a YCSB~\cite{ycsb,ycsbc}, which
emulates real-world key-value store workloads.
We added support for \homa and \sdp to it.
Its default value size is \SI{1}{\KB}, but to see the impact of
sizes, we also test with smaller (\SI{64}{\byte}) or larger (\SI{4}{\KB})
values.
To saturate the server, we use multiple threads and cores at the client, each
opens its own socket to send requests to the server in parallel.

We compare \sdp with or without TLS offload against TCP and TLS/TCP.
Redis uses user-space TLS that does not support
hardware offload, but we added support for kTLS to make a fair
comparison to \sdp.

\sdp outperforms Redis/TLS in all the workloads and value sizes.
\sdp without TLS offload outperforms user-space TLS by 5--24\% and kTLS without
offload by 8--22\%.
When TLS offload is enabled, \sdp outperforms kTLS by 5--18\%.
Recall that the throughput of \homa and \sdp was constrained to around
\SI{700}{K} RPCc/s by the softirq thread in \autoref{sec:eval:concurrent}
(\autoref{fig:multi}).
Since Redis has a considerable amount of application-level processing overheads (e.g.,
request parsing and database manipulation), the overall rates are below that
rate, and thus \homa and \sdp always outperformed the TCP counterparts.

In \SI{64}{B} RPCs, since the application-level processing and data
encryption/transmission happen in the same thread that becomes the bottleneck,
CPU cycles freed up by encryption offload directly improved performance, whereas
at higher request rates with \sdp, the relative cost of the receive path of the
server’s stack became higher and thus we saw a smaller benefit of encryption
offload. 

TCP (without TLS) performs slightly better than \homa with \SI{4}{\KB} items, because it is optimized for large transfers.
However, \sdp, even without TLS offload, always outperforms TLS.
Similar to \SI{8}{\KB} RPC cases in \autoref{fig:multi}, this highlights the
better processing locality achieved by transport-level integration of
cryptographic operations.

\subsection{In-Kernel Client: NVMe-oF}

To show the applicability of \sdp to in-kernel applications, we added
experimental support for it in NVMe-oF.
NVMe-oF is a remote block storage service to connect fast NVMe-based
SSD devices and network clients.
It is implemented in the kernel to avoid moving the data between the user and
kernel spaces. 
If the NVMe-OF stack was implemented in user space, to handle a read
request, the data needs to be moved out of the kernel block layer and then
moving the data back into the kernel to send it over TCP.

Similar to Redis, the most effort was about supporting \homa, and once it was
done, \sdp support was trivial.
It was slightly harder than Redis due to the lack of \homa APIs
for kernel clients, which we thus implemented.
We needed to modify the NVMe-oF layer, because it expects stream
abstraction of the network I/O, whereas \sdp inherits \homa's RPC
abstraction.
Since the NVMe stack is implemented inside the Linux kernel block layer, we can
use unmodified client applications on top of it.

Our current implementation is in early stage and still expensive, including
one extra data copy compared to TCP and lack of support for multiple I/O
queues.

\bh{Results}
We use FIO~\cite{fio}, a widely-used storage benchmark tool, to generate random
read requests to the remote SSD node over TCP, \ktls, \homa or \sdp.
We use the default NVMe block size, \SI{4}{\KB}, and force the data to be read
from the NVMe SSD, not from the page cache.

\begin{figure}[tb]
\centering
	\includegraphics[width=\linewidth]{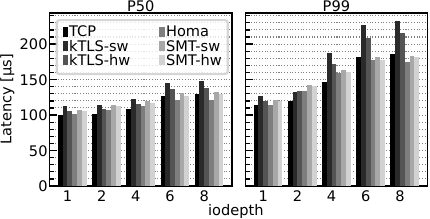}
	\caption{P50 and P99 latency of NVMe-oF.}
\label{fig:nvme}
\end{figure}
\autoref{fig:nvme} plots the P50 and P99 request latency over varying
\textsf{iodepth}, the number of requests sent without waiting for the response
to the previous requests.
We ran each \textsf{iodepth} 5 times, 30 seconds each, and plot the middle one
in P50 latency.
Although we were not able to observe the advantage of \homa or \sdp
when \textsf{iodepth} is 1--4 at P50 or 1--2 at P99, we saw up to \SI{7}{\%}
(with TLS offload) or \SI{15}{\%} (without it) of P50 latency reduction, and up to
\SI{16}{\%} (with TLS offload) or \SI{21}{\%} (without it) of P99 latency
reduction.

Unlike Redis cases, we were not able to observe clear advantage of hardware TLS
offloading, likely because the benefit was masked by other NVMe device or stack
overheads that increase the end-to-end latency.
We leave further analysis and improvement of NVMe-oF/\sdp as future work.

\subsection{Comparison with TCPLS (and QUIC)}
\begin{figure*}[tb]
         \begin{minipage}[b]{0.32\linewidth}
         \includegraphics[width=\linewidth]{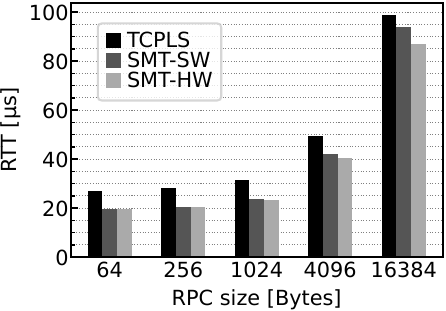}
         \caption{TCPLS comparison.}\label{fig:tcpls}
         \end{minipage}
         \hspace{1mm}
         \begin{minipage}[b]{0.32\linewidth}
         \includegraphics[width=\linewidth]{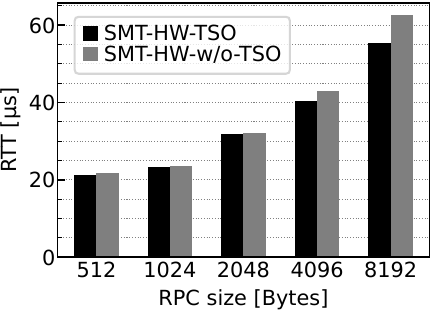}
         \caption{Effect of TSO.}\label{fig:gso}
         \end{minipage}
         \hspace{1mm}
         \begin{minipage}[b]{0.32\linewidth}
         \includegraphics[width=\linewidth]{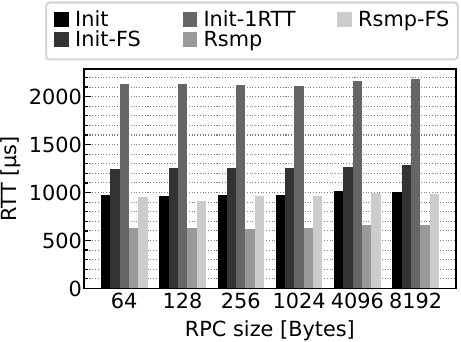}
         \caption{Key exchange latency.}\label{fig:sdpkx}
         \end{minipage}
\end{figure*}
TCPLS~\cite{tcpls} augments TCP by extending TLS 1.3 to achieve the similar
features to QUIC (\autoref{sec:options:enc}).
We compare \sdp with TCPLS, because it outperforms all the QUIC implementations
they tested, including Quicly (the fastest one), Msquic and mvfst, by at least
2.4$\times$~\cite{tcpls}.
\autoref{fig:tcpls} plots unloaded latency to highlight software overheads.
\sdp without TLS offload exhibits \SIrange{5}{18}{\%} lower latency than TCPLS.
\sdp with TLS offload achieves \SIrange{12}{18}{\%} lower latency than TCPLS
(cannot use offload, see \autoref{sec:options:enc}).

\subsection{Key Exchange Performance}
We implemented handshake methods that support 0-RTT data with and without forward
secrecy (\autoref{sec:design:key}). \autoref{fig:sdpkx} shows the RTT of the initial
handshake and session resumption for each method, compared to the baseline, which
performs a standard TLS 1.3 handshake over \homa (without pre-key generation). We use ECDH
key exchange with \textsf{secp256r1}, the \textsf{aes128gcmsha256}
cipher suite, and ECDSA with the \textsf{secp256r1} signature algorithm.

The \sdp initial handshake outperforms standard TLS (Init-1RTT) by
37--44\% when forward secrecy is enabled (Init-FS), otherwise (Init) \SIrange{52}{55}{\%}.
In addition to RTT saving, it eliminates C1.1 in \autoref{table:tlsbreakdown} through
key pre-generation, and C3.1 and C3.2 by verifying the certificate from \sdp-ticket in advance
 on the client side. On the server side, it removes S2.1 by using a pre-generated 
ephemeral key share.

For resumption (denoted as Rsmp), our implementation also uses pre-generated keys at both ends. The
margin between Rsmp-FS and Rsmp (no forward secrecy) is 338--387\us; it is
reasonable, because the additional costs of S2.2 and C2.2 are similar.

\section{Security Analysis}\label{sec:sec}

Security properties of \sdp are based on TLS 1.3~\cite{rfc8446}.
This section first details how those properties are achieved in \sdp design
in \autoref{sec:tls}.
We then discuss attacks outside the explicit defense scope of TLS 1.3
in \autoref{sec:sec:notls}.
Finally, we compare \sdp with other TLS-based transport-level encryptions
in terms of security properties in \autoref{sec:other}.

\subsection{TLS Security Properties}\label{sec:tls}

RFC8446~\cite{rfc8446}, which describes the TLS 1.3 specification, defines the
following security properties in Section 1: \emph{authentication},
\emph{confidentiality}, and \emph{integrity}.
\sdp ensures the authentication properties using TLS 1.3 handshake and thus
inherits protection against attacks that affected earlier TLS versions
(e.g., POODLE~\cite{poodle}, BEAST~\cite{beast}, and Lucky 13~\cite{lucky}).
\sdp ensures confidentiality and integrity using AEAD encryption, which provides
both the properties.

RFC8446 defines additional guarantees that enhance confidentiality and
integrity properties of TLS 1.3 in Section E.2: \emph{order protection},
\emph{non-replayability}, \emph{length concealment}, and \emph{forward secrecy after key change}.
Since \sdp needs to ensure some of those guarantees or mechanisms differently
than TLS/TCP to support message-based transport, we discuss those in detail.

\bh{Order protection}
This property ensures the attacker cannot make the receiver accept the
out-of-order record sequence number.
In TLS/TCP, the (single) byte stream in the connection is
mapped to the TLS session with a single record sequence number space.
Therefore, out-of-order data does not reach the TLS layer and the ordering guarantee of TLS prevents the altered TLS records that
preserve the TCP-level correctness (i.e., sequence number, length and checksum)
from being accepted at the TLS layer.
To support the message abstraction that ensures reliable, in-order byte
delivery \emph{within} the message but different messages can be reordered, \sdp
applies the order protection in a per-message basis using per-message record
sequence number space in a secure session (\autoref{sec:design}).
Within the record sequence number space, the order and completeness of records
are guaranteed by monotonically incrementing record sequence numbers, like
TLS/TCP.

\bh{Non-replayability}
\sdp ensures this guarantee by composite message identities
(\autoref{sec:design:para}).
Per-message record sequence number space means the \emph{relative} record
sequence number can duplicate across the messages in the TLS session.
To avoid replay, \sdp ensures uniqueness of message ID throughout the
session (\autoref{sec:design:msgid}).
When the receiver detects the message ID already seen previously, it simply
discards it without decryption, much like TCP discards the packet with the
past sequence number.
When the receiver receives a new message ID but with altered payload, it detects
replay or injection when decrypting it, much like TLS receiver does so after
receiving an altered segment but correct at the TCP layer (i.e., sequence
number, length and checksum are correct).

\bh{Length concealment}
TLS applications can conceal the true length of the application data to protect
form side-channel attacks (although AEAD is safe without size changing in terms
of confidentiality).
\sdp is compatible with TLS padding.
Each record can include padding as regular TLS/TCP.
When padding is used, the message length field (\autoref{fig:sdpgso}) should
include padding length to be aligned with the purpose of padding (i.e., hiding
the true data length from the plaintext metadata); the mismatch
between the true application message size and what the header indicate is not a
problem, because the receiver can identify the padding length at the time of
decryption (TLS records are reassembled based on packet IDs and TSO offsets,
which happens prior to message-level reassembly (\autoref{sec:design:enc}).

\bh{Forward secrecy after key exchange}
This guarantee protects the user data from the leakage of server's private
key.
\sdp ensures this guarantee by ephemeral (EC)DHE key exchange enforced by the
TLS 1.3 handshake, except for the (optional) 0-RTT data discussed in
\autoref{sec:design:key}.

\subsection{Metadata and Traffic Analysis Considerations}\label{sec:sec:notls}

As noted in RFC8446, TLS 1.3 does not provide specific guarantees for traffic
analysis attacks, although it provides a \emph{mechanism} to
conceal application data length through padding (see \autoref{sec:tls}).
However, it is known that traffic metadata, such as packet timing, sizes, rate
and bursts, could reveal meaning information from encrypted traffic, such as
(candidate) identifies of website~\cite{juarez2014,smith2021website} and video
streaming~\cite{hasselquist2024raising,silhouette}, often using ML-based
methods~\cite{sirinam2018deep,bhat2019var}.
TLS 1.3 also does not provide defense to side-channel attacks caused by
microarchitectural effects.

\sdp also does not provide protection against those attacks except for the
message length concealment mechanism.
A notable metadata that could be exploited by traffic analysis is plaintext message
ID and length field (\autoref{fig:sdpgso}), which could be more meaningful
compared to the sequence number field in TCP headers.
However, if necessary, that can be obfuscated using TLS 1.3 padding, as
discussed in \autoref{sec:tls}.

It is reported that plaintext TLS 1.3 record header could leak some degree of
information to ``weak'' attackers~\cite{ford2016metadata}.
The relevant concern is that TLS record headers are more easily identifiable
when aligned with packet boundaries.
In TCP, when multiple messages together form TCP segments, the positions of the
record headers are unlikely to align with the beginning of the packet payload,
except for the first one.
\sdp currently places TLS record headers aligned with message boundaries
(\autoref{sec:design:enc}), which would ease the eavesdropper to identify
their positions.
Nevertheless, we do not believe the impact of this alignment is particularly
high in comparison to TLS/TCP, because TCP
applications would use \texttt{TCP\_NODELAY} anyways, which reduces opportunities of the
aforementioned record boundary obfuscation.
We leave the analysis of effectiveness of traffic analysis attacks on \sdp traffic as future work.

\subsection{Other TLS-Based Encryptions}\label{sec:other}
It is worthwhile to discuss how other TLS-based protocols guarantee the TLS 1.3
security properties, particularly for those that \sdp guarantees differently
from TLS/TCP.
DTLS runs on UDP so it is impractical to assume that the underlying transport
results in in-order, complete delivery beyond a single UDP datagram boundary.
For the non-replayability guarantee, the DTLS receiver maintains a sliding window that defines the range of
acceptable record sequence numbers and tracks those already received. This
enables the receiver to uniquely identify valid records and discard any that
either fall outside the window or are detected as replays.
\sdp doesn’t use the sliding window, because Homa provides reliable, ordered
byte delivery (like TCP) within the message. This allows each record sequence
number space to assume the same underlying transport property as TCP.

QUIC does not use a record sequence number mechanism; per-stream ordering is
handled above encryption.
Replay protection is provided by a monotonically increasing packet number
incorporated into the AEAD nonce.
Receivers track processed packet numbers to discard duplicates.
Unlike TLS/TCP, QUIC accepts higher packet numbers even if earlier ones are
missing, tolerating reordering.
Therefore, QUIC's encryption mechanism, as shown in \autoref{fig:rec_seq_nounce},
can be seen as per-packet record sequence number space in contrast to
per-message one in \sdp and per-connection one in TLS/TCP.

Note that header protection encrypts the packet number
to prevent middlebox interference and ossification, not as a replay defense.
Unlike QUIC, \sdp does not enforce header protection, because
middlebox ossification is less relevant in datacenters.
This design preserves the property of message-based transports that enables
per-message load balancing between the host stack CPU cores or multiple paths
(\autoref{sec:discuss})

\section{Discussion}\label{sec:discuss}

\bh{Message integrity}
When TSO is used for a non-TCP protocol, the NIC does not embed a checksum in the overlaid
TCP header field, meaning that checksum offload is impossible.
Because of this, \homa does not guarantee message integrity.
This means that the application must compute the checksum and embed it in their
messages.

\sdp intrinsically obviates this problem, because encrypting and decrypting the
message with TLS ensures integrity is verified. Moreover, the cryptographic
operations can be offloaded to the NIC hardware.

\bh{Segmentation}
IPv6 does not have an equivalent field to the IPv4 IPID, which we currently use for
reassembling the TSO segment (\autoref{sec:design:enc}).
We can still use TLS offload without TSO, which we plot the impact in
\autoref{fig:gso}.
Note that the penalty of disabling TSO in \sdp or \homa would not be as large as
in TCP cases.
This is because, although TSO does both segmentation and checksumming, \homa
does not use the latter or guarantee message integrity (\sdp intrinsically does
it based on message encryption and decryption).

We can use TSO for every pair of packets, as the receiver can reassemble them based on
the presence of the TLS record header. For larger TSO segments, we use GSO to split them
into two-packet-sized TSO segments at the bottom of the stack. These smaller TSO segments
contain incrementing TSO offsets generated by GSO (\autoref{fig:sdpgso}).
If the NIC vendor activates sequence number embedding to non-TCP packets, which
could not require significant changes to the NIC implementation, \sdp will
enable full TSO.

\bh{Receive-side offload}
Receiver-side cryptographic NIC offload is challenging because the NIC does not route
incoming non-TCP packets through its cryptographic engine, although it would be a
straightforward task for the vendor to implement. For example, Pensando Elba features
a P4-based ingress packet processing engine~\cite{bansal2023disaggregating} situated
before the cryptographic engine.
Although we confirmed that sender-side offload accelerates RPC throughput by up to
\SIrange{5}{13}{\%}, receive-side offload would further accelerate it.

\bh{Encryption protocol}
\sdp would support PSP used in Falcon (\autoref{sec:options:enc}) in addition to TLS, but the challenges of encrypting message-based
transport protocols discussed in this paper remain valid unless the NIC explicitly supports
the \sdp packet format. This is because encryption would occur per TSO segment, which
includes framing headers, while PSP uses a single encryption offset.
We plan to explore using PSP instead of TLS once a NIC with PSP offload becomes available.

\bh{In-network compute compatibility}
\sdp is compatible with In-Network Compute (INC) enabled by MTP~\cite{mtp},
unlike other encrypted transports discussed in \autoref{sec:options:enc}.
This is because it leaves message ID and length unencrypted and thus allows a network node to identify message boundaries of
application-level messages with bounded resource usage, for example, for
congestion signalling or load balancing.
\sdp is compatible with packet trimming used by NDP~\cite{ndp} and
UET~\cite{uet}, because trimmed
packets carry useful information (i.e., plaintext transport metadata) for the
receiver to identify the sender demands.

\bh{Hardware-based transport}
\sdp is designed to be used as replacement of TLS/TCP in ordinary cloud
environments that provide virtualized or baremetal instances or networks,
allowing opportunistic use of NIC offload for segmentation and encryption.
If implemented entirely in hardware like Falcon/PSP, \sdp would achieve lower
small RPC latency due to end-to-end SRPT scheduling inherited from Homa; \sdp
has some communication overheads for compatibility with existing offload, like
TCP-structured header, whereas Falcon/PSP avoids those overheads with clean-slate packet format supported by the custom NIC.

\bh{Post-quantum resistance}
\sdp inherits post-quantum resistance of TLS 1.3 using an appropriate handshake that
prevents captured handshakes from being attacked.
The user may opt for longer keys for slightly better quantum resistance.
In this case, the benefit of hardware offload would be larger.
Although we used 128-bit key in this paper (\autoref{sec:eval}), the NIC we used
 also supports 256-bit keys for TLS offload.

\section{Related Work}\label{sec:related}
Much of the related work has been discussed in \autoref{sec:options};
the remaining topics are covered here.

\bh{RDMA network security}
ReDMArK~\cite{redmark} demonstrates packet injection attacks in RDMA networks, highlighting 
the need for encryption, such as IPSec or sRDMA~\cite{srdma}, which mitigates these attacks 
by using symmetric cryptography and embedding MAC in the RDMA header.
sRDMA employs symmetric cryptography for authentication and encryption, and extends the RDMA packet header to embed MAC.
The application establishes RDMA connections (QPs) with the agent in the local SmartNIC whose CPUs perform authentication and encryption jobs, attaching or removing outer headers.
\cite{rdmasecvision} proposes encrypting RDMA packets with DTLS using hardware
acceleration, but \sdp does not opt for DTLS, because it needs TSO and large
message support.

\bh{Key exchange acceleration}
SSLShader~\cite{sslshader} and SmartTLS~\cite{smarttls} accelerate the TLS handshake using GPU and SmartNIC, respectively.
Those can be used for \sdp if key exchange is performed based on TLS, although
we explored lightweight methods based on symmetric keys (\autoref{sec:design:key}).

\bh{Host stack enhancements}
Host stack improvements, such as 
batching~\cite{megapipe}, zero copy~\cite{stackmap}, flexible core 
allocation~\cite{netchannel} and better NIC abstraction~\cite{enso}, complement
\sdp, though TCP-specific optimizations like congestion control~\cite{dctcp} and
handshake improvements~\cite{tfo} are not applicable.
ByteDance has reported their effort of improving \homa for their RPC
traffic~\cite{bytedancehoma}, improving large send performance with pipelining,
congestion control with better RTT measurement, loss detection, and buffer
estimation to coexist with TCP traffic. Those techniques are transparently
applicable to \sdp; they report \homa's throughput is lower than TCP when the
message size is larger than \SI{50}{\KB}, which we also observe similarly in \autoref{sec:eval}.

\bh{Transport protocol design}
It is worth discussing design patterns of transport protocols.
Multipath TCP~\cite{mptcpnsdi} focuses on robustness against middlebox interference prevalent in the Internet~\cite{mptcpnsdi}.
Its compatibility with TSO also enables datacenter usage for large transfers~\cite{mptcpdc}, although sharing the problems with TCP for RPCs (\autoref{sec:options:msg}).
SCTP~\cite{rfc9260} defines its own protocol number, which is viable in its primary target, telecommunication networks, but has resulted in low adoption in the Internet due to middlebox interference.
It is also not datacenter friendly due to high software overheads and protocol complexity.
\sdp's design point is support for existing hardware offload and RPCs without
consideration of middleboxes that block transport protocols other than TCP or
UDP, which are prevalent in the Internet.

\bh{Transport multiplexing}
Aquila~\cite{aquila} and EQDS~\cite{eqds} enable sharing of the same network fabric for all 
host traffic, including TCP and RDMA. EQDS operates as \emph{edge functions}, scheduling 
traffic over UDP using an NDP-derived control loop, while Aquila uses a ToR-in-NIC (TiN) 
chip for hardware-based transport (GNet). \sdp can be multiplexed within these systems, 
providing abstraction and encryption to the application, and can also be used between 
edge functions.

\section{Conclusion}\label{sec:conclusion}

We explored a new design point of secure datacenter transport protocols to
transition from the current TLS/TCP ecosystem for more efficient secure
datacenter networking.
We found that the TLS record protocol can be used with new message-based
transports designed for datacenter RPCs together with existing TLS offload
available in commodity NICs, but doing so required tightly coupling transport protocol and encryption to preserve the security properties of TLS/TCP.

The current \sdp implementation inherits performance issues from
\homa~\cite{bytedancehoma}, but we expect those will be mitigated due to its
active development.

\section*{Acknowledgments}
We are grateful to the anonymous reviewers and shepherd for valuable comments.
We thank John Ousterhout for the discussion on Homa.
We also thank Boris Pismenny for helping us understand TLS offload.
This work was in part supported by EPSRC grant EP/V053418/1, Royal Society
Research Grant, and gift from Google and NetApp.

\printbibliography[heading=bibintoc] 

@inproceedings{dctcp,
author = {Alizadeh, Mohammad and Greenberg, Albert and Maltz, David A. and Padhye, Jitendra and Patel, Parveen and Prabhakar, Balaji and Sengupta, Sudipta and Sridharan, Murari},
title = {Data Center TCP (DCTCP)},
year = {2010},
isbn = {9781450302012},
publisher = {Association for Computing Machinery},
address = {New York, NY, USA},
url = {https://doi.org/10.1145/1851182.1851192},
doi = {10.1145/1851182.1851192},
booktitle = {Proceedings of the ACM SIGCOMM 2010 Conference},
pages = {63–74},
numpages = {12},
location = {New Delhi, India},
series = {SIGCOMM '10}
}

@techreport{ietf-tls-esni-17,
    number =    {draft-ietf-tls-esni-17},
    type =      {Internet-Draft},
    institution =   {Internet Engineering Task Force},
    publisher = {Internet Engineering Task Force},
    note =      {Work in Progress},
    url =       {https://datatracker.ietf.org/doc/draft-ietf-tls-esni/17/},
    author =    {Eric Rescorla and Kazuho Oku and Nick Sullivan and Christopher A. Wood},
    title =     {{TLS Encrypted Client Hello}},
    pagetotal = 48,
    year =      2023,
    month =     oct,
    day =       9,
}

@inproceedings {paste,
	author = {Michio Honda and Giuseppe Lettieri and Lars Eggert and Douglas
		Santry},
	title = {{PASTE}: A Network Programming Interface for Non-Volatile Main
		Memory},
	booktitle = {15th {USENIX} Symposium on Networked Systems Design and
		Implementation ({NSDI} 18)},
	year = {2018},
	isbn = {978-1-939133-01-4},
	address = {Renton, WA},
	pages = {17--33},
	url = {https://www.usenix.org/conference/nsdi18/presentation/honda},
	publisher = {{USENIX} Association},
	month = apr,
}

@inproceedings{megapipe,
 author = {Han, Sangjin and Marshall, Scott and Chun, Byung-Gon and Ratnasamy, Sylvia},
 title = {MegaPipe: A New Programming Interface for Scalable Network I/O},
 booktitle = {Proceedings of the 10th USENIX Conference on Operating Systems Design and Implementation},
 series = {OSDI'12},
 year = 2012,
 isbn = {978-1-931971-96-6},
 location = {Hollywood, CA, USA},
 pages = {135--148},
 numpages = 14,
 url = {http://dl.acm.org/citation.cfm?id=2387880.2387894},
 acmid = 2387894,
 publisher = {USENIX Association},
 address = {Berkeley, CA, USA},
}

@inproceedings{mtcp,
 author = {Jeong, Eun Young and Woo, Shinae and Jamshed, Muhammad and Jeong, Haewon and Ihm, Sunghwan and Han, Dongsu and Park, KyoungSoo},
 title = {mTCP: A Highly Scalable User-level TCP Stack for Multicore Systems},
 booktitle = {Proceedings of the 11th USENIX Conference on Networked Systems Design and Implementation},
 series = {NSDI'14},
 year = 2014,
 isbn = {978-1-931971-09-6},
 location = {Seattle, WA},
 pages = {489--502},
 numpages = 14,
 url = {http://dl.acm.org/citation.cfm?id=2616448.2616493},
 acmid = 2616493,
 publisher = {USENIX Association},
 address = {Berkeley, CA, USA},
}

@inproceedings {stackmap,
	author = {Kenichi Yasukata and Michio Honda and Douglas Santry and Lars Eggert},
	title = {StackMap: Low-Latency Networking with the OS Stack and Dedicated NICs},
	booktitle = {2016 USENIX Annual Technical Conference (USENIX ATC 16)},
	year = {2016},
	isbn = {978-1-931971-30-0},
	address = {Denver, CO},
	pages = {43--56},
	url = {https://www.usenix.org/conference/atc16/technical-sessions/presentation/yasukata},
	publisher = {USENIX Association},
}

@inproceedings {pegasus,
	author = {Jialin Li and Jacob Nelson and Ellis Michael and Xin Jin and
		Dan R. K. Ports},
	title = {Pegasus: Tolerating Skewed Workloads in Distributed Storage
		with In-Network Coherence Directories},
	booktitle = {14th {USENIX} Symposium on Operating Systems Design and
		Implementation ({OSDI} 20)},
	year = {2020},
	isbn = {978-1-939133-19-9},
	pages = {387--406},
	url = {https://www.usenix.org/conference/osdi20/presentation/li-jialin},
	publisher = {{USENIX} Association},
	month = nov,
}

@inproceedings {homalinux,
title = {A Linux Kernel Implementation of the Homa Transport Protocol},
author = {John Ousterhout},
booktitle = {2021 {USENIX} Annual Technical Conference ({USENIX} {ATC} 21)},
year = {2021},
url = {https://www.usenix.org/conference/atc21/presentation/ousterhout},
publisher = {{USENIX} Association},
month = jul,
}

@inproceedings{ndp,
	author = {Handley, Mark and Raiciu, Costin and Agache, Alexandru and
		Voinescu, Andrei and Moore, Andrew W. and Antichi, Gianni and
			W\'{o}jcik, Marcin},
	title = {Re-Architecting Datacenter Networks and Stacks for Low Latency
		and High Performance},
	year = {2017},
	isbn = {9781450346535},
	publisher = {Association for Computing Machinery},
	address = {New York, NY, USA},
	url = {https://doi.org/10.1145/3098822.3098825},
	doi = {10.1145/3098822.3098825},
	booktitle = {Proceedings of the Conference of the ACM Special Interest
		Group on Data Communication},
	pages = {29–42},
	numpages = {14},
	location = {Los Angeles, CA, USA},
	series = {SIGCOMM '17}
}

@inproceedings{homa,
author = {Montazeri, Behnam and Li, Yilong and Alizadeh, Mohammad and Ousterhout, John},
title = {Homa: A Receiver-Driven Low-Latency Transport Protocol Using Network Priorities},
year = {2018},
isbn = {9781450355674},
publisher = {Association for Computing Machinery},
address = {New York, NY, USA},
url = {https://doi.org/10.1145/3230543.3230564},
doi = {10.1145/3230543.3230564},
booktitle = {Proceedings of the 2018 Conference of the ACM Special Interest Group on Data Communication},
pages = {221–235},
numpages = {15},
location = {Budapest, Hungary},
series = {SIGCOMM '18}
}

@misc{rfc9000,
    series =    {Request for Comments},
    number =    9000,
    howpublished =  {RFC 9000},
    publisher = {RFC Editor},
    doi =       {10.17487/RFC9000},
    url =       {https://www.rfc-editor.org/info/rfc9000},
    author =    {Jana Iyengar and Martin Thomson},
    title =     {{QUIC: A UDP-Based Multiplexed and Secure Transport}},
    pagetotal = 151,
    year =      2021,
    month =     may,
}

@misc{rfc9001,
    series =    {Request for Comments},
    number =    9001,
    howpublished =  {RFC 9001},
    publisher = {RFC Editor},
    doi =       {10.17487/RFC9001},
    url =       {https://www.rfc-editor.org/info/rfc9001},
    author =    {Martin Thomson and Sean Turner},
    title =     {{Using TLS to Secure QUIC}},
    pagetotal = 52,
    year =      2021,
    month =     may,
}

@inproceedings{tcpls,
author = {Rochet, Florentin and Assogba, Emery and Piraux, Maxime and Edeline, Korian and Donnet, Benoit and Bonaventure, Olivier},
title = {TCPLS: Modern Transport Services with TCP and TLS},
year = {2021},
isbn = {9781450390989},
publisher = {Association for Computing Machinery},
address = {New York, NY, USA},
url = {https://doi.org/10.1145/3485983.3494865},
doi = {10.1145/3485983.3494865},
booktitle = {Proceedings of the 17th International Conference on Emerging Networking EXperiments and Technologies},
pages = {45–59},
numpages = {15},
location = {Virtual Event, Germany},
series = {CoNEXT '21}
}

@inproceedings{dcquic,
  author={Tan, Lizhuang and Su, Wei and Liu, Yanwen and Gao, Xiaochuan and Zhang, Wei},
  booktitle={IEEE INFOCOM WKSHPS},
  title={DCQUIC: Flexible and Reliable Software-defined Data Center Transport}, 
  year={2021},
  volume={},
  number={},
  pages={1-8},
  doi={10.1109/INFOCOMWKSHPS51825.2021.9484596}}

@inproceedings {mtp,
author = {Tao Ji and Rohan Vardekar and Balajee Vamanan and Brent E. Stephens and Aditya Akella},
title = {{MTP}: Transport for {In-Network} Computing},
booktitle = {22nd USENIX Symposium on Networked Systems Design and Implementation (NSDI 25)},
year = {2025},
isbn = {978-1-939133-46-5},
address = {Philadelphia, PA},
pages = {959--977},
url = {https://www.usenix.org/conference/nsdi25/presentation/ji},
publisher = {USENIX Association},
month = apr
}

@inproceedings{tcpcrypt,
  title={The Case for Ubiquitous Transport-Level Encryption},
  author={Bittau, Andrea and Hamburg, Michael and Handley, Mark and Mazieres, David and Boneh, Dan},
  booktitle={19th USENIX Security Symposium (USENIX Security 10)},
  year={2010}
}

@misc{rfc8548,
    series =    {Request for Comments},
    number =    8548,
    howpublished =  {RFC 8548},
    publisher = {RFC Editor},
    doi =       {10.17487/RFC8548},
    url =       {https://www.rfc-editor.org/info/rfc8548},
    author =    {Andrea Bittau and Daniel B. Giffin and Mark J. Handley and David Mazieres and Quinn Slack and Eric W. Smith},
    title =     {{Cryptographic Protection of TCP Streams (tcpcrypt)}},
    pagetotal = 32,
    year =      2019,
    month =     may,
}

@misc{dcns,
    title = {Data center networking stack},
    year = {2016},
    author = {Tom Herbert},
    howpublished = {The Technical Conference on Linux Networking (Netdev 1.2), \url{https://legacy.netdevconf.info/1.2/session.html?tom-herbert/}},
}

@misc{bytedancehoma,
    title = {Leveraging Homa: Enhancing Datacenter RPC Transport Protocols},
    year = {2023},
    author = {Xiaochun Lu and zijian Zhang},
    howpublished = {The Technical Conference on Linux Networking (Netdev 0x17), \url{https://netdevconf.info/0x17/docs/netdev-0x17-paper36-talk-paper.pdf}},
}

@misc{homarecvbuf,
    title = {Kernel-Managed User Buffers in Homa},
    year = {2023},
    author = {John Ousterhout},
    howpublished = {The Technical Conference on Linux Networking (Netdev 0x17),
	    \url{https://netdevconf.info/0x17/docs/netdev-0x17-paper38-talk-slides/HomaBuffersNetDev.pdf}},
}

@misc{enctransit,
    title = {Encryption in transit for Google Cloud},
    year = {2025 (last updated)},
    author = {Google},
    howpublished = {\url{https://cloud.google.com/docs/security/encryption-in-transit}},
}

@misc{encfb,
    title = {Building Facebook’s service encryption infrastructure},
    year = {2019},
    author = {Meta},
    howpublished = {\url{https://engineering.fb.com/2019/05/29/security/service-encryption/}},
}

@misc{pqfb,
    title = {Post-quantum readiness for TLS at Meta},
    year = {2024},
    author = {Meta},
    howpublished = {\url{https://engineering.fb.com/2024/05/22/security/post-quantum-readiness-tls-pqr-meta/}},
}

@misc{cloudflaretls,
    title = {An overview of TLS 1.3 and Q\&A},
    year = {2016},
    author = {Filippo Valsorda},
    howpublished = {\url{https://blog.cloudflare.com/tls-1-3-overview-and-q-and-a}},
}

@misc{davieeastwest,
    title = {The Challenge of East-West Traffic},
    year = {2023},
    author = {Bruce Davie},
    howpublished = {\url{https://systemsapproach.substack.com/p/the-challenge-of-east-west-traffic}},
}

@inproceedings{exttcp,
author = {Honda, Michio and Nishida, Yoshifumi and Raiciu, Costin and Greenhalgh, Adam and Handley, Mark and Tokuda, Hideyuki},
title = {Is It Still Possible to Extend TCP?},
year = {2011},
isbn = {9781450310130},
publisher = {Association for Computing Machinery},
address = {New York, NY, USA},
url = {https://doi.org/10.1145/2068816.2068834},
doi = {10.1145/2068816.2068834},
booktitle = {Proceedings of the 2011 ACM SIGCOMM Conference on Internet Measurement Conference},
pages = {181–194},
numpages = {14},
location = {Berlin, Germany},
series = {IMC '11}
}

@misc{psp,
    title = {Announcing PSP's cryptographic hardware offload at scale is now open source},
    author = {Google},
    howpublished = {\url{https://cloud.google.com/blog/products/identity-security/announcing-psp-security-protocol-is-now-open-source}},
}

@misc{ntflxtls,
    title = {Serving Netflix Video Traffic at 400Gb/s and Beyond},
    author = {Drew Gallatin},
    howpublished = {\url{https://nabstreamingsummit.com/wp-content/uploads/2022/05/2022-Streaming-Summit-Netflix.pdf}},
}

@inproceedings{rktio,
author = {Thalheim, J\"{o}rg and Unnibhavi, Harshavardhan and Priebe, Christian and Bhatotia, Pramod and Pietzuch, Peter},
title = {Rkt-Io: A Direct I/O Stack for Shielded Execution},
year = {2021},
isbn = {9781450383349},
publisher = {Association for Computing Machinery},
address = {New York, NY, USA},
url = {https://doi.org/10.1145/3447786.3456255},
doi = {10.1145/3447786.3456255},
booktitle = {Proceedings of the Sixteenth European Conference on Computer Systems},
pages = {490–506},
numpages = {17},
location = {Online Event, United Kingdom},
series = {EuroSys '21}
}

@misc{ycsbc,
    title={YCSB-C},
    author = {Jinglei Ren},
    howpublished = {\url{https://github.com/basicthinker/YCSB-C}},
}

@misc{kcm,
    title = {Kernel Connection Multiplexer},
    howpublished = {\url{https://www.kernel.org/doc/Documentation/networking/kcm.txt}},
}

@misc{cilium,
    title = {eBPF-based Networking, Security, and Observability},
    author = {Cilium},
    howpublished = {\url{https://github.com/cilium/cilium}},
}

@misc{ciliummtls,
    title = {Improving the security of Cilium Mutual Authentication},
    author = {Cilium},
    howpublished = {\url{https://cilium.io/blog/2024/03/20/improving-mutual-auth-security/}}
}

@inproceedings {eqds,
author = {Vladimir Olteanu and Haggai Eran and Dragos Dumitrescu and Adrian Popa and Cristi Baciu and Mark Silberstein and Georgios Nikolaidis and Mark Handley and Costin Raiciu},
title = {An edge-queued datagram service for all datacenter traffic},
booktitle = {19th USENIX Symposium on Networked Systems Design and Implementation (NSDI 22)},
year = {2022},
isbn = {978-1-939133-27-4},
address = {Renton, WA},
pages = {761--777},
url = {https://www.usenix.org/conference/nsdi22/presentation/olteanu},
publisher = {USENIX Association},
month = apr,
}

@inproceedings {aquila,
author = {Dan Gibson and Hema Hariharan and Eric Lance and Moray McLaren and Behnam Montazeri and Arjun Singh and Stephen Wang and Hassan M. G. Wassel and Zhehua Wu and Sunghwan Yoo and Raghuraman Balasubramanian and Prashant Chandra and Michael Cutforth and Peter Cuy and David Decotigny and Rakesh Gautam and Alex Iriza and Milo M. K. Martin and Rick Roy and Zuowei Shen and Ming Tan and Ye Tang and Monica Wong-Chan and Joe Zbiciak and Amin Vahdat},
title = {Aquila: A unified, low-latency fabric for datacenter networks},
booktitle = {19th USENIX Symposium on Networked Systems Design and Implementation (NSDI 22)},
year = {2022},
isbn = {978-1-939133-27-4},
address = {Renton, WA},
pages = {1249--1266},
url = {https://www.usenix.org/conference/nsdi22/presentation/gibson},
publisher = {USENIX Association},
month = apr,
}

@inproceedings{autonomous,
  title={Autonomous NIC offloads},
  author={Pismenny, Boris and Eran, Haggai and Yehezkel, Aviad and Liss, Liran and Morrison, Adam and Tsafrir, Dan},
  booktitle={Proceedings of the 26th ACM International Conference on Architectural Support for Programming Languages and Operating Systems},
  pages={18--35},
  year={2021}
}

@inproceedings{seemakhupt2023cloud,
  title={A Cloud-Scale Characterization of Remote Procedure Calls},
  author={Seemakhupt, Korakit and Stephens, Brent E and Khan, Samira and Liu, Sihang and Wassel, Hassan and Yeganeh, Soheil Hassas and Snoeren, Alex C and Krishnamurthy, Arvind and Culler, David E and Levy, Henry M},
  booktitle={Proceedings of the 29th Symposium on Operating Systems Principles},
  pages={498--514},
  year={2023}
}

@inproceedings{minion,
  title={Fitting Square Pegs Through Round Pipes: Unordered Delivery Wire-Compatible with TCP and TLS},
  author={Nowlan, Michael F and Tiwari, Nabin and Iyengar, Janardhan and Amin, Syed Obaid and Ford, Bryan},
  booktitle={9th USENIX Symposium on Networked Systems Design and Implementation (NSDI 12)},
  pages={383--398},
  year={2012}
}

@inproceedings{maglev,
  title={Maglev: A fast and reliable software network load balancer},
  author={Eisenbud, Danielle E and Yi, Cheng and Contavalli, Carlo and Smith, Cody and Kononov, Roman and Mann-Hielscher, Eric and Cilingiroglu, Ardas and Cheyney, Bin and Shang, Wentao and Hosein, Jinnah Dylan},
  booktitle={13th USENIX Symposium on Networked Systems Design and Implementation (NSDI 16)},
  pages={523--535},
  year={2016}
}

@inproceedings{snap,
  title={Snap: A microkernel approach to host networking},
  author={Marty, Michael and de Kruijf, Marc and Adriaens, Jacob and Alfeld, Christopher and Bauer, Sean and Contavalli, Carlo and Dalton, Michael and Dukkipati, Nandita and Evans, William C and Gribble, Steve and others},
  booktitle={Proceedings of the 27th ACM Symposium on Operating Systems Principles},
  pages={399--413},
  year={2019}
}

@misc{rfc8446,
    series =    {Request for Comments},
    number =    8446,
    howpublished =  {RFC 8446},
    publisher = {RFC Editor},
    doi =       {10.17487/RFC8446},
    url =       {https://www.rfc-editor.org/info/rfc8446},
    author =    {Eric Rescorla},
    title =     {{The Transport Layer Security (TLS) Protocol Version 1.3}},
    pagetotal = 160,
    year =      2018,
    month =     aug,
}

@misc{ktlsdoc,
    title = {Kernel TLS offload},
    howpublished = {\url{https://www.kernel.org/doc/html/latest/networking/tls-offload.html}}
}

@inproceedings{netchannel,
  title={Towards $\mu$ s tail latency and terabit ethernet: disaggregating the host network stack},
  author={Cai, Qizhe and Vuppalapati, Midhul and Hwang, Jaehyun and Kozyrakis, Christos and Agarwal, Rachit},
  booktitle={Proceedings of the ACM SIGCOMM 2022 Conference},
  pages={767--779},
  year={2022}
}

@inproceedings{tfo,
  title={TCP fast open},
  author={Radhakrishnan, Sivasankar and Cheng, Yuchung and Chu, Jerry and Jain, Arvind and Raghavan, Barath},
  booktitle={Proceedings of the Seventh COnference on emerging Networking EXperiments and Technologies},
  pages={1--12},
  year={2011}
}

@inproceedings{rdmasecvision,
  title={Securing RDMA for High-Performance Datacenter Storage Systems},
  author={Simpson, Anna Kornfeld and Szekeres, Adriana and Nelson, Jacob and Zhang, Irene},
  booktitle={12th USENIX Workshop on Hot Topics in Cloud Computing (HotCloud 20)},
  year={2020}
}

@inproceedings{smarttls,
  title={A case for smartnic-accelerated private communication},
  author={Kim, Duckwoo and Lee, SeungEon and Park, KyoungSoo},
  booktitle={4th Asia-Pacific Workshop on Networking},
  pages={30--35},
  year={2020}
}

@inproceedings{sslshader,
  title={SSLShader: Cheap SSL Acceleration with Commodity Processors},
  author={Jang, Keon and Han, Sangjin and Han, Seungyeop and Moon, Sue and Park, KyoungSoo},
  booktitle={8th USENIX Symposium on Networked Systems Design and Implementation (NSDI 11)},
  year={2011}
}

@inproceedings {redmark,
author = {Benjamin Rothenberger and Konstantin Taranov and Adrian Perrig and Torsten Hoefler},
title = {{ReDMArk}: Bypassing {RDMA} Security Mechanisms},
booktitle = {30th USENIX Security Symposium (USENIX Security 21)},
year = {2021},
isbn = {978-1-939133-24-3},
pages = {4277--4292},
url = {https://www.usenix.org/conference/usenixsecurity21/presentation/rothenberger},
publisher = {USENIX Association},
month = aug
}

@inproceedings{srdma,
  title={sRDMA--Efficient NIC-based Authentication and Encryption for Remote Direct Memory Access},
  author={Taranov, Konstantin and Rothenberger, Benjamin and Perrig, Adrian and Hoefler, Torsten},
  booktitle={2020 USENIX Annual Technical Conference (USENIX ATC 20)},
  pages={691--704},
  year={2020}
}

@inproceedings{ford2016metadata,
  title={Metadata Protection Considerations for TLS Present and Future},
  author={Ford, Bryan Alexander},
  booktitle={TLS 1.3 Ready or Not (TRON) Workshop},
  year={2016}
}

@inproceedings{bansal2023disaggregating,
  title={Disaggregating Stateful Network Functions},
  author={Bansal, Deepak and DeGrace, Gerald and Tewari, Rishabh and Zygmunt, Michal and Grantham, James and Gai, Silvano and Baldi, Mario and Doddapaneni, Krishna and Selvarajan, Arun and Arumugam, Arunkumar and others},
  booktitle={20th USENIX Symposium on Networked Systems Design and Implementation (NSDI 23)},
  pages={1469--1487},
  year={2023}
}

@inproceedings{ycsb,
  title={Benchmarking cloud serving systems with YCSB},
  author={Cooper, Brian F and Silberstein, Adam and Tam, Erwin and Ramakrishnan, Raghu and Sears, Russell},
  booktitle={Proceedings of the 1st ACM Symposium on Cloud Computing},
  pages={143--154},
  year={2010}
}

@inproceedings{acceltcp,
  title={AccelTCP: Accelerating network applications with stateful TCP offloading},
  author={Moon, YoungGyoun and Lee, SeungEon and Jamshed, Muhammad Asim and Park, KyoungSoo},
  booktitle={17th USENIX Symposium on Networked Systems Design and Implementation (NSDI 20)},
  pages={77--92},
  year={2020}
}

@inproceedings{demikernel,
  title={The demikernel datapath os architecture for microsecond-scale datacenter systems},
  author={Zhang, Irene and Raybuck, Amanda and Patel, Pratyush and Olynyk, Kirk and Nelson, Jacob and Leija, Omar S Navarro and Martinez, Ashlie and Liu, Jing and Simpson, Anna Kornfeld and Jayakar, Sujay and others},
  booktitle={Proceedings of the ACM SIGOPS 28th Symposium on Operating Systems Principles},
  pages={195--211},
  year={2021}
}

@techreport{piraux-tcpls-00,
    number =    {draft-piraux-tcpls-00},
    type =      {Internet-Draft},
    institution =   {Internet Engineering Task Force},
    publisher = {Internet Engineering Task Force},
    note =      {Work in Progress},
    url =       {https://datatracker.ietf.org/doc/draft-piraux-tcpls/00/},
    author =    {Maxime Piraux and Olivier Bonaventure and Florentin Rochet},
    title =     {{TCPLS: Modern Transport Services with TCP and TLS}},
    pagetotal = 19,
    year =      2021,
    month =     oct,
    day =       25,
}

@techreport{davie-stt-05,
    number =    {draft-davie-stt-05},
    type =      {Internet-Draft},
    institution =   {Internet Engineering Task Force},
    publisher = {Internet Engineering Task Force},
    note =      {Work in Progress},
    url =       {https://datatracker.ietf.org/doc/draft-davie-stt/05/},
    author =    {Bruce Davie and Jesse Gross},
    title =     {{A Stateless Transport Tunneling Protocol for Network Virtualization (STT)}},
    pagetotal = 20,
    year =      2014,
    month =     mar,
    day =       13,
}

@misc{ktlscilium,
    title = {Seamless transparent encryption with BPF and Cilium},
    author = {John Fastabend},
    howpublished = {Linux Plumbers Conference 2019},
}

@misc{ktlscilium2,
    title = {Combining kTLS and BPF for Introspection and Policy Enforcement},
    author = {Daniel Borkmann and John Fastabend},
    howpublished = {Linux Plumbers Conference 2018},
}

@inproceedings{falcon,
author = {Singhvi, Arjun and Dukkipati, Nandita and Chandra, Prashant and Wassel, Hassan M. G. and Sharma, Naveen Kr. and Rebello, Anthony and Schuh, Henry and Kumar, Praveen and Montazeri, Behnam and Bansod, Neelesh and Thomas, Sarin and Cho, Inho and Seibert, Hyojeong Lee and Wu, Baijun and Yang, Rui and Li, Yuliang and Huang, Kai and Yin, Qianwen and Agarwal, Abhishek and Vaduvatha, Srinivas and Wang, Weihuang and Moshref, Masoud and Ji, Tao and Wetherall, David and Vahdat, Amin},
title = {Falcon: A Reliable, Low Latency Hardware Transport},
year = {2025},
isbn = {9798400715242},
publisher = {Association for Computing Machinery},
address = {New York, NY, USA},
url = {https://doi.org/10.1145/3718958.3754353},
doi = {10.1145/3718958.3754353},
booktitle = {Proceedings of the ACM SIGCOMM 2025 Conference},
pages = {248–263},
numpages = {16},
location = {S\~{a}o Francisco Convent, Coimbra, Portugal},
series = {SIGCOMM '25}
}

@inproceedings{mptcpnsdi,
  title={How hard can it be? designing and implementing a deployable multipath TCP},
  author={Raiciu, Costin and Paasch, Christoph and Barre, Sebastien and Ford, Alan and Honda, Michio and Duchene, Fabien and Bonaventure, Olivier and Handley, Mark},
  booktitle={9th USENIX symposium on Networked Systems Design and Implementation (NSDI 12)},
  pages={399--412},
  year={2012}
}

@inproceedings{mptcpdc,
author = {Raiciu, Costin and Barre, Sebastien and Pluntke, Christopher and Greenhalgh, Adam and Wischik, Damon and Handley, Mark},
title = {Improving datacenter performance and robustness with multipath TCP},
year = {2011},
isbn = {9781450307970},
publisher = {Association for Computing Machinery},
address = {New York, NY, USA},
url = {https://doi.org/10.1145/2018436.2018467},
doi = {10.1145/2018436.2018467},
booktitle = {Proceedings of the ACM SIGCOMM 2011 Conference},
pages = {266–277},
numpages = {12},
location = {Toronto, Ontario, Canada},
series = {SIGCOMM '11}
}

@misc{rfc9260,
    series =    {Request for Comments},
    number =    9260,
    howpublished =  {RFC 9260},
    publisher = {RFC Editor},
    doi =       {10.17487/RFC9260},
    url =       {https://www.rfc-editor.org/info/rfc9260},
    author =    {Randall R. Stewart and Michael Tüxen and karen Nielsen},
    title =     {{Stream Control Transmission Protocol}},
    pagetotal = 133,
    year =      2022,
    month =     jun,
}

@misc{nvidiashare,
    title = {Mellanox Corporate Update---Unleashing the Power of Data},
    year = {2020},
    author = {Mellanox Technologies},
}

@inproceedings{nicmem,
  title={The benefits of general-purpose on-NIC memory},
  author={Pismenny, Boris and Liss, Liran and Morrison, Adam and Tsafrir, Dan},
  booktitle={Proceedings of the 27th ACM International Conference on Architectural Support for Programming Languages and Operating Systems},
  pages={1130--1147},
  year={2022}
}

@article{srd,
  title={A cloud-optimized transport protocol for elastic and scalable hpc},
  author={Shalev, Leah and Ayoub, Hani and Bshara, Nafea and Sabbag, Erez},
  journal={IEEE micro},
  volume={40},
  number={6},
  pages={67--73},
  year={2020},
  publisher={IEEE}
}

@misc{ubertcp,
    title = {Better Load Balancing: Real-Time Dynamic Subsetting},
    author = {Chien-Chih Liao and Pawel Krolikowski, Sangeeta Kundu},
    howpublished = {\url{https://www.uber.com/en-GB/blog/better-load-balancing-real-time-dynamic-subsetting/}},
}

@inproceedings {crisp,
author = {Zhizhou Zhang and Murali Krishna Ramanathan and Prithvi Raj and Abhishek Parwal and Timothy Sherwood and Milind Chabbi},
title = {{CRISP}: Critical Path Analysis of {Large-Scale} Microservice Architectures},
booktitle = {2022 USENIX Annual Technical Conference (USENIX ATC 22)},
year = {2022},
isbn = {978-1-939133-29-50},
address = {Carlsbad, CA},
pages = {655--672},
url = {https://www.usenix.org/conference/atc22/presentation/zhang-zhizhou},
publisher = {USENIX Association},
month = jul
}

@inproceedings {servicerouter,
author = {Harshit Saokar and Soteris Demetriou and Nick Magerko and Max Kontorovich and Josh Kirstein and Margot Leibold and Dimitrios Skarlatos and Hitesh Khandelwal and Chunqiang Tang},
title = {{ServiceRouter}: Hyperscale and Minimal Cost Service Mesh at Meta},
booktitle = {17th USENIX Symposium on Operating Systems Design and Implementation (OSDI 23)},
year = {2023},
isbn = {978-1-939133-34-2},
address = {Boston, MA},
pages = {969--985},
url = {https://www.usenix.org/conference/osdi23/presentation/saokar},
publisher = {USENIX Association},
month = jul
}

@inproceedings {tectonic,
author = {Satadru Pan and Theano Stavrinos and Yunqiao Zhang and Atul Sikaria and Pavel Zakharov and Abhinav Sharma and Shiva Shankar P and Mike Shuey and Richard Wareing and Monika Gangapuram and Guanglei Cao and Christian Preseau and Pratap Singh and Kestutis Patiejunas and JR Tipton and Ethan Katz-Bassett and Wyatt Lloyd},
title = {Facebook{\textquoteright}s Tectonic Filesystem: Efficiency from Exascale},
booktitle = {19th USENIX Conference on File and Storage Technologies (FAST 21)},
year = {2021},
isbn = {978-1-939133-20-5},
pages = {217--231},
url = {https://www.usenix.org/conference/fast21/presentation/pan},
publisher = {USENIX Association},
month = feb
}

@inproceedings{rpcbench,
author = {Gan, Yu and Zhang, Yanqi and Cheng, Dailun and Shetty, Ankitha and Rathi, Priyal and Katarki, Nayan and Bruno, Ariana and Hu, Justin and Ritchken, Brian and Jackson, Brendon and Hu, Kelvin and Pancholi, Meghna and He, Yuan and Clancy, Brett and Colen, Chris and Wen, Fukang and Leung, Catherine and Wang, Siyuan and Zaruvinsky, Leon and Espinosa, Mateo and Lin, Rick and Liu, Zhongling and Padilla, Jake and Delimitrou, Christina},
title = {An Open-Source Benchmark Suite for Microservices and Their Hardware-Software Implications for Cloud \& Edge Systems},
year = {2019},
isbn = {9781450362405},
publisher = {Association for Computing Machinery},
address = {New York, NY, USA},
url = {https://doi.org/10.1145/3297858.3304013},
doi = {10.1145/3297858.3304013},
booktitle = {Proceedings of the Twenty-Fourth International Conference on Architectural Support for Programming Languages and Operating Systems},
pages = {3–18},
numpages = {16},
location = {Providence, RI, USA},
series = {ASPLOS '19}
}

@inproceedings{dirigent,
author = {Cvetkovi\'{c}, Lazar and Costa, Fran\c{c}ois and Djokic, Mihajlo and Friedman, Michal and Klimovic, Ana},
title = {Dirigent: Lightweight Serverless Orchestration},
year = {2024},
isbn = {9798400712517},
publisher = {Association for Computing Machinery},
address = {New York, NY, USA},
url = {https://doi.org/10.1145/3694715.3695966},
doi = {10.1145/3694715.3695966},
booktitle = {Proceedings of the ACM SIGOPS 30th Symposium on Operating Systems Principles},
pages = {369–384},
numpages = {16},
location = {Austin, TX, USA},
series = {SOSP '24}
}

@inproceedings {appnet,
author = {Xiangfeng Zhu and Yuyao Wang and Banruo Liu and Yongtong Wu and Nikola Bojanic and Jingrong Chen and Gilbert Louis Bernstein and Arvind Krishnamurthy and Sam Kumar and Ratul Mahajan and Danyang Zhuo},
title = {High-level Programming for Application Networks},
booktitle = {22nd USENIX Symposium on Networked Systems Design and Implementation (NSDI 25)},
year = {2025},
isbn = {978-1-939133-46-5},
address = {Philadelphia, PA},
pages = {915--935},
url = {https://www.usenix.org/conference/nsdi25/presentation/zhu},
publisher = {USENIX Association},
month = apr
}

@inproceedings{quicnotfast,
author = {Zhang, Xumiao and Jin, Shuowei and He, Yi and Hassan, Ahmad and Mao, Z. Morley and Qian, Feng and Zhang, Zhi-Li},
title = {QUIC is not Quick Enough over Fast Internet},
year = {2024},
isbn = {9798400701719},
publisher = {Association for Computing Machinery},
address = {New York, NY, USA},
url = {https://doi.org/10.1145/3589334.3645323},
doi = {10.1145/3589334.3645323},
booktitle = {Proceedings of the ACM Web Conference 2024},
pages = {2713–2722},
numpages = {10},
location = {Singapore, Singapore},
series = {WWW '24}
}

@misc{homaup,
    title = {Begin upstreaming Homa transport protocol},
    author = {John Ousterhout},
    howpublished = {\url{https://lwn.net/Articles/997858/}},
}

@misc{psplinux,
    title = {[RFC net-next 00/15] add basic PSP encryption for TCP connections},
    author = {Jakub Kicinski},
    year = {2024},
    month = may,
    howpublished = {\url{https://lore.kernel.org/all/20240510030435.120935-1-kuba@kernel.org/}},
}

@inproceedings{fastconfidential,
author = {Lefeuvre, Hugo and Chisnall, David and Kogias, Marios and Olivier, Pierre},
title = {Towards (Really) Safe and Fast Confidential I/O},
year = {2023},
isbn = {9798400701955},
publisher = {Association for Computing Machinery},
address = {New York, NY, USA},
url = {https://doi.org/10.1145/3593856.3595913},
doi = {10.1145/3593856.3595913},
booktitle = {Proceedings of the 19th Workshop on Hot Topics in Operating Systems},
pages = {214–222},
numpages = {9},
location = {Providence, RI, USA},
series = {HOTOS '23}
}

@inproceedings {scone,
author = {Sergei Arnautov and Bohdan Trach and Franz Gregor and Thomas Knauth and Andre Martin and Christian Priebe and Joshua Lind and Divya Muthukumaran and Dan O{\textquoteright}Keeffe and Mark L. Stillwell and David Goltzsche and Dave Eyers and R{\"u}diger Kapitza and Peter Pietzuch and Christof Fetzer},
title = {{SCONE}: Secure Linux Containers with Intel {SGX}},
booktitle = {12th USENIX Symposium on Operating Systems Design and Implementation (OSDI 16)},
year = {2016},
isbn = {978-1-931971-33-1},
address = {Savannah, GA},
pages = {689--703},
url = {https://www.usenix.org/conference/osdi16/technical-sessions/presentation/arnautov},
publisher = {USENIX Association},
month = nov
}

@article{poodle,
  title={This POODLE bites: exploiting the SSL 3.0 fallback},
  author={M{\"o}ller, Bodo and Duong, Thai and Kotowicz, Krzysztof},
  journal={Security Advisory},
  volume={21},
  pages={34--58},
  year={2014}
}

@article{beast,
  title={Here Come The $\oplus$ Ninjas},
  author={Duong, Thai and Rizzo, Juliano},
  journal={Unpublished manuscript},
  volume={320},
  year={2011}
}

@inproceedings{lucky,
  title={Lucky thirteen: Breaking the TLS and DTLS record protocols},
  author={Al Fardan, Nadhem J and Paterson, Kenneth G},
  booktitle={IEEE S\&P},
  pages={526--540},
  year={2013},
}

@inproceedings{sirinam2018deep,
	title        = {Deep Fingerprinting: Undermining Website Fingerprinting Defenses With Deep Learning},
	author       = {Sirinam, Payap and Imani, Mohsen and Juarez, Marc and others},
	year         = 2018,
	booktitle    = CCS,
	publisher    = {{ACM}},
	pages        = {1928--1943}
}

@article{hasselquist2024raising,
  title={Raising the Bar: Improved Fingerprinting Attacks and Defenses for Video Streaming Traffic},
  author={Hasselquist, David and Witwer, Ethan and Carlson, August and Johansson, Niklas and Carlsson, Niklas},
  journal={Proceedings on Privacy Enhancing Technologies},
  year={2024}
}

@inproceedings{juarez2014,
author = {Juarez, Marc and Afroz, Sadia and Acar, Gunes and Diaz, Claudia and Greenstadt, Rachel},
title = {A Critical Evaluation of Website Fingerprinting Attacks},
year = {2014},
isbn = {9781450329576},
publisher = {Association for Computing Machinery},
address = {New York, NY, USA},
url = {https://doi.org/10.1145/2660267.2660368},
doi = {10.1145/2660267.2660368},
booktitle = {Proceedings of the 2014 ACM SIGSAC Conference on Computer and Communications Security},
pages = {263–274},
numpages = {12},
location = {Scottsdale, Arizona, USA},
series = {CCS '14}
}

@inproceedings{silhouette,
  title={Silhouette: Identifying youtube video flows from encrypted traffic},
  author={Li, Feng and Chung, Jae Won and Claypool, Mark},
  booktitle={Proceedings of the 28th acm sigmm workshop on network and operating systems support for digital audio and video},
  pages={19--24},
  year={2018}
}

@article{smith2021website,
  title={Website Fingerprinting in the Age of QUIC},
  author={Smith, J and Mittal, P and Perrig, A},
  journal={Proceedings on Privacy Enhancing Technologies},
  year={2021}
}

@article{bhat2019var,
  title={Var-CNN: A Data-Efficient Website Fingerprinting Attack Based on Deep Learning},
  author={Bhat, Sanjit and Lu, David and Kwon, Albert and Devadas, Srinivas},
  journal={Proceedings on Privacy Enhancing Technologies},
  year={2019}
}

@misc{gellman2013nsa,
  title={NSA infiltrates links to Yahoo, Google data centers worldwide, Snowden documents say},
  author={Gellman, Barton and Soltani, Ashkan},
  howpublished={The Washington Post, \url{https://www.washingtonpost.com/world/national-security/nsa-infiltrates-links-to-yahoo-google-data-centers-worldwide-snowden-documents-say/2013/10/30/e51d661e-4166-11e3-8b74-d89d714ca4dd_story.html}},
  year={2013}
}

@misc{equifax,
  title={2017 Equifax data breach},
  howpublished={\url{https://en.wikipedia.org/wiki/2017_Equifax_data_breach}},
  year={2017}
}

@inproceedings{privateeye,
  title={PrivateEye: Scalable and Privacy-Preserving compromise detection in the cloud},
  author={Arzani, Behnaz and Ciraci, Selim and Saroiu, Stefan and Wolman, Alec and Stokes, Jack and Outhred, Geoff and Diwu, Lechao},
  booktitle={17th USENIX Symposium on Networked Systems Design and Implementation (NSDI 20)},
  pages={797--815},
  year={2020}
}

@inproceedings{mogulphysical,
author = {Mogul, Jeffrey C. and Wilkes, John},
title = {Physical Deployability Matters},
year = {2023},
isbn = {9798400704154},
publisher = {Association for Computing Machinery},
address = {New York, NY, USA},
url = {https://doi.org/10.1145/3626111.3628190},
doi = {10.1145/3626111.3628190},
booktitle = {Proceedings of the 22nd ACM Workshop on Hot Topics in Networks},
pages = {9–17},
numpages = {9},
location = {Cambridge, MA, USA},
series = {HotNets '23}
}

@inproceedings {aura,
author = {Sivaramakrishnan Ramanathan and Ying Zhang and Mohab Gawish and Yogesh Mundada and Zhaodong Wang and Sangki Yun and Eric Lippert and Walid Taha and Minlan Yu and Jelena Mirkovic},
title = {Practical Intent-driven Routing Configuration Synthesis},
booktitle = {20th USENIX Symposium on Networked Systems Design and Implementation (NSDI 23)},
year = {2023},
isbn = {978-1-939133-33-5},
address = {Boston, MA},
pages = {629--644},
url = {https://www.usenix.org/conference/nsdi23/presentation/ramanathan},
publisher = {USENIX Association},
month = apr
}

@inproceedings {netvigil,
author = {Kevin Hsieh and Mike Wong and Santiago Segarra and Sathiya Kumaran Mani and Trevor Eberl and Anatoliy Panasyuk and Ravi Netravali and Ranveer Chandra and Srikanth Kandula},
title = {{NetVigil}: Robust and {Low-Cost} Anomaly Detection for {East-West} Data Center Security},
booktitle = {21st USENIX Symposium on Networked Systems Design and Implementation (NSDI 24)},
year = {2024},
isbn = {978-1-939133-39-7},
address = {Santa Clara, CA},
pages = {1771--1789},
url = {https://www.usenix.org/conference/nsdi24/presentation/hsieh},
publisher = {USENIX Association},
month = apr
}

@inproceedings {openbypass,
  author = {Shinichi Awamoto and Michio Honda},
  title = {Opening Up Kernel-Bypass TCP Stacks},
  booktitle = {2025 USENIX Annual Technical Conference (USENIX ATC)},
  pages = {249–262},
  year = {2025},
  url = {https://www.usenix.org/conference/atc25/presentation/awamoto},
  publisher = {{USENIX} Association},
  month = jul,
}

@inproceedings{hatonen2010experimental,
  title={An experimental study of home gateway characteristics},
  author={H{\"a}t{\"o}nen, Seppo and Nyrhinen, Aki and Eggert, Lars and Strowes, Stephen and Sarolahti, Pasi and Kojo, Markku},
  booktitle={Proceedings of the 10th ACM SIGCOMM conference on Internet measurement},
  pages={260--266},
  year={2010}
}

@inproceedings{edeline2019bottom,
  title={A bottom-up investigation of the transport-layer ossification},
  author={Edeline, Korian and Donnet, Benoit},
  booktitle={2019 Network Traffic Measurement and Analysis Conference (TMA)},
  pages={169--176},
  year={2019},
  publisher={IEEE}
}

@inproceedings {enso,
author = {Hugo Sadok and Nirav Atre and Zhipeng Zhao and Daniel S. Berger and James C. Hoe and Aurojit Panda and Justine Sherry and Ren Wang},
title = {{Enso}: A Streaming Interface for {NIC-Application} Communication},
booktitle = {17th USENIX Symposium on Operating Systems Design and Implementation (OSDI 23)},
year = {2023},
isbn = {978-1-939133-34-2},
address = {Boston, MA},
pages = {1005--1025},
url = {https://www.usenix.org/conference/osdi23/presentation/sadok},
publisher = {USENIX Association},
month = jul
}

@misc{uet,
    title = {Ultra Ethernet Specification v1.0.1},
    author = {Ultra Ethernet Consortium},
    year={2025},
    month={Sept},
    howpublished = {\url{https://ultraethernet.org/wp-content/uploads/sites/20/2025/10/UE-Specification-1.0.1.pdf}},
}

@inproceedings {eng25519,
author = {Jipeng Zhang and Junhao Huang and Lirui Zhao and Donglong Chen and {\c C}etin Kaya Ko{\c c}},
title = {{ENG25519}: Faster {TLS} 1.3 handshake using optimized X25519 and Ed25519},
booktitle = {33rd USENIX Security Symposium (USENIX Security 24)},
year = {2024},
isbn = {978-1-939133-44-1},
address = {Philadelphia, PA},
pages = {6381--6398},
url = {https://www.usenix.org/conference/usenixsecurity24/presentation/zhang-jipeng},
publisher = {USENIX Association},
month = aug
}

@misc{fio,
    title = {Flexible IO Tester (FIO)},
    author = {Jens Axboe},
    howpublished = {\url{https: //github.com/axboe/fio}}
}

\end{document}